%% file: manuscript.tex
\documentclass{IEEEtranModified}
\usepackage{schemabloc}
\usepackage[cmex10]{amsmath}
\usepackage{amsthm}
\usepackage{amsfonts}
\usepackage{amssymb}
\usepackage{footnote}
\usepackage{cite}
\usepackage{graphicx}
\usepackage{epstopdf}
\usepackage{setspace}
\usepackage{lineno}
\usepackage{url}
\usepackage{shadethm}
\usepackage{color, colortbl}
\usepackage[caption=false]{subfig}
\usepackage{lineno}
\usepackage{array}
\usepackage{xspace}
\usepackage[latin1]{inputenc}
\usepackage{tikz}
\usetikzlibrary{chains,scopes}
\usetikzlibrary{shapes,arrows}
\usepackage{stackrel}
\usepackage{mathtools}

\newcommand{\diag}{\mathsf{diag}}

\newcommand{\comps}{\mathbb{C}}

\newcommand{\Mysubfloat}[4]{
    \subfloat[#1]{\includegraphics[width=#2\textwidth]{#3}\label{#4}}}
\DeclareGraphicsExtensions{.eps,.jpg,.pdf,.png}
\makeatletter
\let\old@ps@headings\ps@headings
\let\old@ps@IEEEtitlepagestyle\ps@IEEEtitlepagestyle
\def\articlefooter#1{%
    \def\ps@headings{%
        \old@ps@headings%
        \def\@oddfoot{\strut\hfill#1\hfill\strut}%
        \def\@evenfoot{\strut\hfill#1\hfill\strut}%
    }%
    \def\ps@IEEEtitlepagestyle{%
        \old@ps@IEEEtitlepagestyle%
        \def\@oddfoot{\strut\hfill#1\hfill\strut}%
        \def\@evenfoot{\strut\hfill#1\hfill\strut}%
    }%
    \ps@headings%
}
\makeatother
\articlefooter{%
        \parbox{\textwidth}{\hrulefill \\ \small{Preprint. Work in progress.} \hfill \begin{minipage}{0.2\textwidth}\centering \vspace*{4pt}  \end{minipage} \hfill \small{ }}%
}

\newcommand{\andr}[1]{{{\textcolor{red}{Andrey: #1}}}}

\newcommand{\obs}{\mathcal{O}}
\newcommand{\numpow}{N_s}
\newcommand{\numvol}{N_v}
\newcommand{\ourdsse}{SFSE($T, t, \numpow(t), \numvol(t)$)}

\begin{document}
%%%%%%%%%%%% Framed Specifications Environment%%%%%%%%%%%%%%%
\newshadetheorem{boxdef}{Specification}
\newshadetheorem{boxtheorem}[boxdef]{Theorem*}
\newtheorem{theorem}[boxdef]{Theorem*}
\setlength{\shadeboxsep}{2pt}
\setlength{\shadeboxrule}{.4pt}
\setlength{\shadedtextwidth}{0.49\textwidth}
\addtolength{\shadedtextwidth}{-2\shadeboxsep}
\addtolength{\shadedtextwidth}{-2\shadeboxrule}
\setlength{\shadeleftshift}{0pt}
\setlength{\shaderightshift}{0pt}
\definecolor{shadethmcolor}{cmyk}{0,0,0,0}
\definecolor{shaderulecolor}{cmyk}{0,0,0,1}

\title{Physics-Informed Deep Neural Network Method for Limited Observability  State Estimation}

\author{
\IEEEauthorblockN{Jonatan Ostrometzky$^*$, Konstantin Berestizshevsky$^\dagger$}
\IEEEauthorblockA{* Dept. of Electrical Engineering \\
Columbia University,
New York, NY, USA\\
jonatan.o@columbia.edu; konsta9@mail.tau.ac.il \\ \vspace*{-1.5cm}}
\and
\IEEEauthorblockN{Andrey Bernstein$^\ddagger$, Gil Zussman$^*$}
\IEEEauthorblockA{$\dagger$ School of EE, Tel Aviv University, Tel Aviv, Israel \\
$\ddagger$ National Renewable Energy Laboratory, Golden, CO, USA\\
andrey.bernstein@nrel.gov; gil.zussman@columbia.edu \\ \vspace*{-1.5cm}}}

% make the title area
\maketitle

\begin{abstract}
The precise knowledge regarding the state of the power grid is important in
order to ensure optimal and reliable grid operation. Specifically, knowing the
state of the distribution grid becomes increasingly important as more renewable
energy sources are connected directly into the distribution network, increasing
the fluctuations of the injected power.

In this paper, we consider the case when the distribution grid becomes
partially observable, and the state estimation problem is under-determined.
We present a new methodology that leverages a deep neural network (DNN) to
estimate the grid state. The standard DNN training method is modified to
explicitly incorporate the physical information of the grid topology and
line/shunt admittance. We show that our method leads to a superior accuracy
of the estimation when compared to the case when no physical information is
provided. Finally, we compare the performance of our method to the standard
state estimation approach, which is based on the weighted least squares with
pseudo-measurements, and show that our method performs significantly better
with respect to the estimation accuracy.
\end{abstract}


\thanksto{J. Ostrometzky and K. Berestizshevsky contributed equally to this work.
 This work was authored in part by the National Renewable Energy Laboratory
 (NREL), operated by Alliance for Sustainable Energy, LLC, for the U.S.
 Department of Energy (DOE) under Contract No. DE-AC36-08GO28308. The views
 expressed in the article do not necessarily represent the views of the DOE or
 the U.S. Government. The U.S. Government retains and the publisher, by
 accepting the article for publication, acknowledges that the U.S. Government
 retains a nonexclusive, paid-up, irrevocable, worldwide license to publish or
 reproduce the published form of this work, or allow others to do so, for U.S.
 Government purposes. This work was supported in part by the Laboratory
 Directed Research and Development (LDRD) Program at NREL,  U.S. DOE OE as part
 of the DOE Grid Modernization Initiative, U.S. DOE Energy Efficiency and
 Renewable Energy Solar Energy Technologies Office,  DARPA RADICS under
 Contract FA-8750- 16-C-0054, and DTRA grant HDTRA1-13-1-0021.}

\section{Introduction}

The problem of \emph{state estimation} (SE) in power grids involves the
estimation of a part of the variables of the power-flow equations (PFE), based
on noisy measurements of the available variables. The SE is an essential
inference tool that enables advanced control and automation capabilities for
the grid operator. Examples of these capabilities are the Volt/Var control
during a normal operation and feeder reconfiguration during restoration from an
emergency. In the fully observable case, namely, when the number of unknown
variables is smaller than the number of the available measurements, the
weighted least squares (WLS) method \cite{Abur2004} is  well developed and
widely used by the utilities. In this paper, we focus on the
\emph{under-determined} case, namely when the number of the unknowns is larger
than the number of measurements. This is a typical case in distribution
networks, where the measurement infrastructure is
insufficient~\cite{baran2001challenges}, as well as in grids under
failures/attacks.

\begin{figure}[t!]
  \centering
  \input{HighLevel.tex}
  \caption{A high level illustration of the proposed DNN-based distributed system state estimation. The estimation process  infers the voltages phasors $\underline{\hat{v}}(t)$ at the partially observable (PO) time step $t$, based on the previous $T-1$ fully observable time steps. All the measurements at time-step $t$ are used during training as the ground truth (GT). The red parts depict the training procedure which enforces the power flow equations feasibility.}\label{fig:highlevel}
\end{figure}
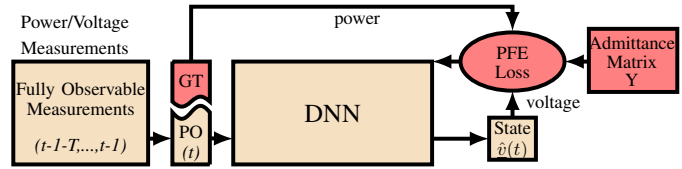

The challenge of identifying a critical failure/attack  and recovering the
network state was the subject of many studies, with a major focus on
transmission systems
\cite{soltan2018react,soltan2019expose,soltan2016quantifying}. As for the
distribution system state estimation (DSSE) problem, it becomes increasingly
important due to the extensive penetration of highly-fluctuating
power-injection sources such as renewable energy sources (e.g., photo-voltaic
panels \cite{ostrometzky2019irradiance}). Thus, DSSE has been studied
extensively in the recent literature; for review of the state-of-the-art
methods, see, e.g.,  \cite{primadianto2017review,wang2019distribution} and
references therein. Sensor placement strategies have been devised to derive
additional measurements required for full observability \cite{singh2009,
bhela2018, YCNN}. Machine learning (ML) and signal processing tools have been
used to derive \emph{pseudo-measurements} using existing sensors and/or
historical data \cite{Clements2011,manitsas2012, wu2013}, and to use them for
further estimation purposes. However, installing sensors might be prohibitively
costly, whereas off-the-shelf ML methods require large amounts of data to
obtain a good estimation accuracy. Finally, several recent works attempt to
directly address the under-determined estimation problem by leveraging the
\emph{low-rank structure} of the measurements
\cite{gao_wang2016,Genes2019,Liao2019,Schmitt}. These approaches can be viewed
as \emph{regularized} WLS methods, wherein the regularization term is aimed at
minimizing the rank of the data matrix.

In this paper, we consider the DSSE problem, in which the grid is observable
during normal operation, and becomes unobservable unexpectedly due to, e.g.,
cyber attacks or physical failures. A high-level diagram of our methodology is
depicted in {\textbf{Fig. \ref{fig:highlevel}}}. Our methodology leverages a
deep neural network (DNN) to estimate the grid state based on historical data.
Unlike standard DNN training methods, here, we modify the training to
explicitly incorporate the physical information of the grid topology based on
the line/shunt admittance. This is achieved by using the AC PFE as a \emph{loss
function regularizer} during the training of the DNN. This modification allows
to reduce the size of the DNN solution search space, which in turn contributes
to a better convergence during the training of the model, and leads to a higher
estimation accuracy compared to the standard (non-regularized) methods.

Our approach can be viewed as a \emph{physics-informed} ML approach that
leverages the physical structure of the problem to improve the performance of a
standard ML algorithm. In this respect, our work is closely related to
\cite{zamzam2019physics}, where the authors proposed to use the approximate
separability property of the DSSE, and designed a pruned DNN via placement of
phasor measurements units (PMUs) at key points in the grid. Our approach, on
the other hand, optimizes the SE task by exploiting the known admittance matrix
information, without specific PMU placement requirements, and can be applied to
the existing grid infrastructure. The main contribution of our paper is an
improved DSSE methodology under limited observability, combining a DNN with a
physics-informed regularization.

%%%%%%%%%%%%%%%%%%%%
To demonstrate the proposed method, we employ an experimental setup based on
the IEEE 37-Node test feeder \cite{schneider2018ieee37} with real-world
generation and load data. We show the performance of our method on a number of
partially-observable scenarios, achieving a consistently more accurate
estimations, when compared to: (i) DNN-based approaches that do not use the PFE
information; and (ii) the standard WLS method, based upon basic
pseudo-measurements (which replaces the unobservable power phasors).

The rest of this paper is organized as follows: In Section II we present the
definitions, the theory, and the development of our approach. Section III
describes the experiments and the analysis of the resulting estimates. Lastly,
Section IV concludes this paper and discusses future research challenges.

\section*{Nomenclature}
\begin{table}[htbp!]
\centering
\begin{tabular}{c|c|l}
   Term  & Domain & Description \\
   \hline
    $T$    & $\mathbb{N}$ & Num. of observation time steps \\
    $N$    & $\mathbb{N}$ & Num. of nodes in the grid \\
    $\numpow(t)$ & $\mathbb{N}$    & Num. of observable power phasors \\
    $\numvol(t)$ & $\mathbb{N}$    & Num. of observable voltage phasors \\
    $Y$    & $\mathbb{C}^{N\times N}$ & Admittance matrix \\
    $\underline{s}(t)$    & $\mathbb{C}^{N}$ & Vector of complex power phasors\\
    $\underline{v}(t)$    & $\mathbb{C}^{N}$ & Vector of complex voltage phasors \\
    $\obs^t$ & $\mathbb{R}$ & Degree of observability; $;\in[0,1]$ \\
    $\lambda$ & $\mathbb{R}$ & Regularization coefficient; $\in[0,\infty)$

\end{tabular}
\label{tab:nomenclature}
\end{table}

\section{Theory and Methodology}

In order to present our methodology, we first need to define the environment of interest, and to formulate the problem.

\subsection{Physical Environmental Assumptions}\label{sec:assumptions}
We assume a known grid topology and sensing capabilities, detailed as follows:
\begin{enumerate}
    \item The distribution grid topology is fixed and the admittance matrix, $Y$, is considered known.
    \item Each bus in the grid can report at each time step one of the
        following types of measurements: (i) none, (ii) power phasor, and
        (iii) power and voltage phasors.
    \item During normal operation the grid is fully observable, that is both
        the power and voltage phasors are reported from all the buses at
        regular times. Prior to a partially observable time step, the
        estimator has guaranteed access to $T-1$ fully observable time steps.

    \item For brevity, we focus on a single-phase estimation, basing our
        development on the single-phase AC PFE. The extension to multi-phase
        networks is straightforward, by leveraging multi-phase PFE as in,
        e.g., \cite{linearLoadFlow}.
\end{enumerate}

\subsection{Power System Model and Observability} \label{sec:observability}
We consider a  distribution network consisting of one slack bus and $N-1$
PQ-buses. The PFE are given by
\begin{align} \label{PFE}
\underline{s}(t) = \diag(\underline{v}(t)) Y^*\underline{v}^*(t)
\end{align}
where $\underline{s}(t) \in \comps^N; \underline{v}(t) \in \comps^N$ are
complex vectors that collect the apparent power injections and voltage phasors
at time index $t$, respectively.

We assume that each bus can report at a given time step, $t$, one of the
following combinations of measurements:  $\{\emptyset; s(t) ; \{s(t),v(t)\}\}$,
where $s(t)$ is the apparent power injection, and $v(t)$ is the phasor of the
complex bus voltage. $\emptyset$ represents the empty set, meaning, that no
measurement is being reported at time step $t$ by the bus. Let $N_s(t)$ and
$N_v(t)$ denote the number of the observable power phasors and voltage phasors,
respectively, at time step $t$.

For any $N_s(t)\in \{0, \ldots, N\}$; $N_v(t) \in \{0, \ldots, N\}$, let
\begin{equation}  \label{eq:O}
    \left.\obs (N_s(t),N_v(t)) \triangleq \frac{N_s(t)+N_v(t)}{2N}\right.
\end{equation}
denote  the \emph{degree of observability} of the system. \iffalse Based on the
available measurements, let $\obs^t(N_s,N_v)$ represent the \emph{Degree of
Observability} for a system with $N$ buses for a given time step $t$, in
following manner:
\begin{align}
    \obs^t(N_s,N_v) \triangleq & \begin{cases}
        \frac{N_s}{2N}, & \mbox{if }N_s < N \mbox{~\& }N_v=0 \\
        \frac{1}{2}, & \mbox{if }N_s = N \mbox{~\& }N_v=0 \\
        \frac{1}{2}+\frac{N_v}{2N}, & \mbox{if }N_s = N \mbox{~\& } N_v>0 \\
        1, & \mbox{if }N_v = N \mbox{~\& }0\leq{N_s}\leq{N} \\
        undefined, & \mbox{otherwise}
        \end{cases}
      \end{align}
Where $N_s$ is the number of the observable power phasors, and $N_v$ is the
number of the observable voltage phasors. \andr{Might need to modify the next
text based on the def above.} \fi We also use the shorthand notation $\obs^t :=
\obs (N_s(t),N_v(t))$ to denote the degree of observability at time $t$. Note
that $\obs^t=1$ represents an over-determined system (at time step $t$),
whereas $\obs^t=0$ represents the zero-observability case (in which all buses
report $\emptyset$). Also, observe that $\obs^t \geq 0.5$ is a necessary
condition for solvability of (1) at time $t$.
%Note - there is a problem with the label for some unknown reason...
% So I hardcoded the (1)... if we change the order, we need to manually
% change this (or if we have time to solve the label problem....

\subsection{Problem Formulation}\label{sec:problem}
This section defines the specification of the DSSE problem according to the
assumptions described in Section~\ref{sec:assumptions}. Let us denote by a
\emph{Sudden-Failure-State-Estimator} (SFSE) an estimator that respects the
following specification:

\begin{boxdef}
\textbf{\ourdsse}
\begin{description}\label{spec:sfse}
    \item[Inputs: ]~\\
    \emph{ full history before time $t$:} $\text{ }\{\underline{s}(\tau),\underline{v}(\tau)\}_{\tau=t-1-T}^{t-1}$\\
        \emph{partial~measurements~set~at~time~$t$:~~~~~~~~} \\ $[s_1(t),\ldots,s_{\numpow(t)}(t)];[v_1(t),\ldots,v_{\numvol(t)}(t)]$.\\
    \item[Output: ] ~\\
    $\emph{ voltage estimation at time $t$:} \text{ }\underline{\hat{v}}(t)$.

\end{description}
 \end{boxdef}
Here $T$ is a design parameter that denotes the guaranteed number of time steps
that are fully observable ($N_s(\tau)=N_v(\tau)=N$, and therefore
$\obs^{\tau}=1; \tau = t - 1 -T, \ldots, t-1$), prior to the time of the
failure $t$, at which $\obs^t < 0.5$. In other words, given a fully observable
history of $T-1$ time steps, the goal is to complete the non-observable portion
of the measurements in the last time step.

\subsection{DNN-Based SFSE Solution}\label{sec:nn}
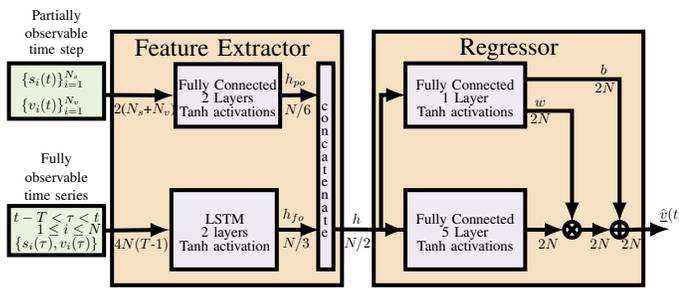
\begin{figure}[htbp]
      \centering
      \input{NN_Architecture.tex}
      \caption{DNN model for the DSSE problem under the SFSE specifications.}\label{fig:nn2}
    \end{figure}

    We propose a neural-network model to solve the \ourdsse\ problem (as
    formulated in Section \ref{sec:problem}). This model is depicted as a block
    diagram in {\textbf{Fig.~\ref{fig:nn2}}}, and consists of two main
    components: \emph{feature extractor} and \textit{regressor}.

    The \textit{feature extractor} receives as an input the
    standardized\footnote{Each data point was reduced by its mean and divided
    by its standard deviation. The mean and the standard deviation were
    computed based on the training set.} phasor data of the fully observable
    period $[t-T-1,t-1]$, and of the partially observable time step $t$. The
    phasors are assumed to be in a rectangular representation, but to simplify
    the architecture and the back-propagation process, we avoid using complex
    numbers in the neural network. Thus, each complex input term is treated as
    two real numbers (based on the real and imaginary parts of the original
    complex term). In essence, each observable time step is represented by a
    $4N$-dimensional vector containing $N$ pairs that represent the power
    phasors ($\underline{s}(\tau)$) and $N$ pairs that represent the voltage
    phasors ($\underline{v}(\tau)$). The feature extractor employs
    Long-Short-Term-Memory (LSTM) \cite{gers1999learning} module to extract
    temporal features $\underline{h}_{fo} \in
    \mathbb{R}^{\lceil{}\frac{N}{3}\rceil{}}$ from the fully observable time
    series. In addition, a fully connected neural network computes the features
    $\underline{h}_{po} \in \mathbb{R}^{\lceil{}\frac{N}{6}\rceil{}}$ from the
    partially observable time step. The $\underline{h}_{fo}$ and the
    $\underline{h}_{po}$ features are finally concatenated to form a unified
    feature vector $\underline{h} \in
    \mathbb{R}^{\lceil{}\frac{N}{2}\rceil{}}$. The sizes of the feature vectors
    were found empirically based on a smaller data set.

    The \textit{regressor} component of the model infers from the feature
    vector $\underline{h}$ the full voltage phasor vector,
    $\underline{\hat{v}}(t)$, represented by a $2N$ dimensional real-valued
    vector, where each pair of values corresponds to a voltage phasor in a
    rectangular form. The regressor module employs fully connected neural
    networks to find the non-scaled estimate of the voltage phasors and to
    determine the coefficients $\underline{w},\underline{b}\in\mathbb{R}^{2N}$.
    These coefficients scale and shift the non-scaled estimates to obtain the
    final voltage estimation, $\underline{\hat{v}}(t)$.

    In the proposed DNN (depicted in {\textbf{Fig. \ref{fig:nn2}}}) all the
  fully connected networks~\cite{nielsen2015neural} use hyperbolic tangent activation functions to
  allow both positive and negative ranges. In the \emph{feature extractor},
  the fully connected network is of the shape $2(\numpow +
  \numvol)\rightarrow2(\numpow + \numvol)\rightarrow N/6$. The LSTM module is
  composed of two stacked LSTM, where the
  first takes sequence of $4N$ long vectors and produces a sequence of
  vectors of size $2N$ which are fed to the second LSTM that produces a
  single output vector of a size $N/3$. In the
  \emph{regressor}, the top fully connected network is in fact two fully
  connected sub-networks in parallel, each taking a vector of a size
  $\frac{N}{2}$ and producing a $2N$-sized vector. The bottom fully
  connected network is of the shape $\frac{N}{2}\rightarrow
  \frac{N}{2}\rightarrow \frac{N}{2}\rightarrow \frac{N}{2}\rightarrow
  \frac{N}{2}\rightarrow 2N$.

\subsection{Physics-Informed Loss Function}\label{sec:loss}

To design a physics-informed training algorithm, we introduce a regularization
term into the DNN loss function $\mathcal{L}$, formalized as
follows\footnote{For better readability, we omit the time indexes from the
following equation, while in fact the time index is $t$.}:
\begin{IEEEeqnarray}{cCl} \label{eqloss}
     \mathcal{L}(\underline{s},\underline{v},\underline{\hat{v}},Y,\lambda)=    ||\underline{v} - \underline{\hat{v}}||^2 + {\lambda}||\underline{s}-\diag(\underline{\hat{v}}) Y^*\underline{\hat{v}}^*||^2 \IEEEyesnumber
\end{IEEEeqnarray}
where the vectors $\underline{s}(t), \underline{v}(t)\in\mathbb{C}^N$ contain
the entries of the measured complex power and voltage variables, respectively;
the vector $\underline{\hat{v}}(t)\in\mathbb{C}^N$ contains the DNN estimated
output of the complex voltage; and $\lambda\in\mathbb{R}$ is a weighting
coefficient. The $||\underline{z}||$ operation over the complex vector
$\underline{z}\in\mathbb{C}^N$ is defined by $\sum_{i=1}^{N}|z_i|^2$. Note that
the vector $\underline{s}(t)$ contains all $N$ power phasors, whereas only
$\numpow(t)$ elements are introduced to the DNN as an input.

In essence, \eqref{eqloss} expresses a loss function comprised of two terms.
The first term is a standard squared-error expression, whereas the second term
penalizes non-feasible PFE solutions. The hyperparameter $\lambda$ is used to
set the ratio between the two terms.

\subsection{Weighted Least Squares Baseline}\label{sec:wls}
 This section describes the WLS-based solution of the SFSE problem as stated in
 section~\ref{sec:problem}. At the time of the failure, $t$, we assume an input
 consisting of previous $T-1$ fully observable time steps, containing $N$ power
 phasors and $N$ voltage phasors. At time $t$, only $\numpow(t)$ power phasors
 and $\numvol(t)$ voltage phasors are available. We mostly focus on the case
 where $\numvol(t)=0$ (all the voltages phasors are unobservable) and we
 perform several experiments using different $\numpow(t)$ values. The WLS
 method will be used as a baseline for comparison against our DNN-based
 approach.

 As a preliminary stage, the missing $N-\numpow(t)$ power phasors are completed
 by duplicating  their last known values (from time step $t-1$). The weights
 for the WLS optimization are intended to re-weight the PFE residuals. These
 weights are computed as follows:
 \begin{align}
     W_i\triangleq
     1/\text{std}\Big(\{s_i(\tau)\}_{\tau=t-1}^{t-1-T}\Big)
 \end{align}
for $1\le i\le N$, where $\text{std}(\{s_i(\tau)\})$ is the standard deviation
of the complex sequence $\{s_i(\tau)\}$, given by the sum of the standard
deviation of the real parts and the imaginary parts of the sequence. The
essence of this re-weighting is to give higher weights to measurements which
are less likely to fluctuate (i.e., have a lower temporal variance).

The optimization variables are the voltage phasors
($\underline{\hat{v}}(t)=[\hat{v}_1(t),\ldots,\hat{v}_N(t)]$) in a rectangular
representation. The initial guesses for the terms of $\underline{\hat{v}}(t)$
are set to $\underline{v}(t-1)$.

With $\numpow(t)$ observable powers and $N-\numpow(t)$ speculatively completed
power phasors gathered in a vector $\hat{s}(t)$ and with the admittance matrix
$Y$ fully known, the WLS optimizer solves the following minimization
problem\textsuperscript{2}:
 \begin{align}
 \min_{\underline{\hat{v}}} F(\underline{\hat{v}}) := \frac{1}{2}\sum_{i=1}^{N}W_i( \Re\{f_{Y,\underline{\hat{s}}}(\underline{\hat{v}},i)\}^2+\Im\{f_{Y,\underline{\hat{s}}}(\underline{\hat{v}},i)\}^2)
 \end{align}
where the function
$f_{Y,\underline{\hat{s}}(t)}(\underline{\hat{v}}(t),i):\mathbb{C}^N\rightarrow\mathbb{C}$
is the residual function obtained from the difference between the
left-hand-side and the right-hand-side of the $i$th PFE with respect to the
admittance matrix $Y\in\mathbb{C}^{N\times N}$ and the power and voltage
complex vectors $\underline{\hat{s}}(t),\underline{v}(t)\in\mathbb{C}^N$.
Namely\textsuperscript{2}:
\begin{align}\label{eqn:residual_function}
    f_{Y,\underline{\hat{s}}}(\underline{\hat{v}},i)\triangleq\hat{s}_i-\sum_{j=1}^{N}\hat{v}_i\cdot Y_{i,j}\cdot \hat{v}_j^*
\end{align}
The optimization procedure for this problem followed the Levenberg-Marquardt
method using the implementation available via the \emph{scipy} Python module.
\section{Experimental Demonstration}
In order to demonstrate our methodology, we designed a dedicated experimental
setup, based on real-world data.
\subsection{Experimental Design and Data Preparation}
Our experimental setup is based on a real-world distribution-grid load data
collected from feeders in Anatolia, CA during the week of August 2012
\cite{bank2013development}. The data includes the active power consumed (i.e.,
the active load) of eight houses, sampled continuously at a 1-second resolution
for a full week (overall, $604800$ samples per house). In addition, the
generation power from a PV panel-array that was connected to the same feeder
was also recorded (at the same time indexes).

\begin{figure}[!ht]
\centering
\includegraphics[width=90mm]{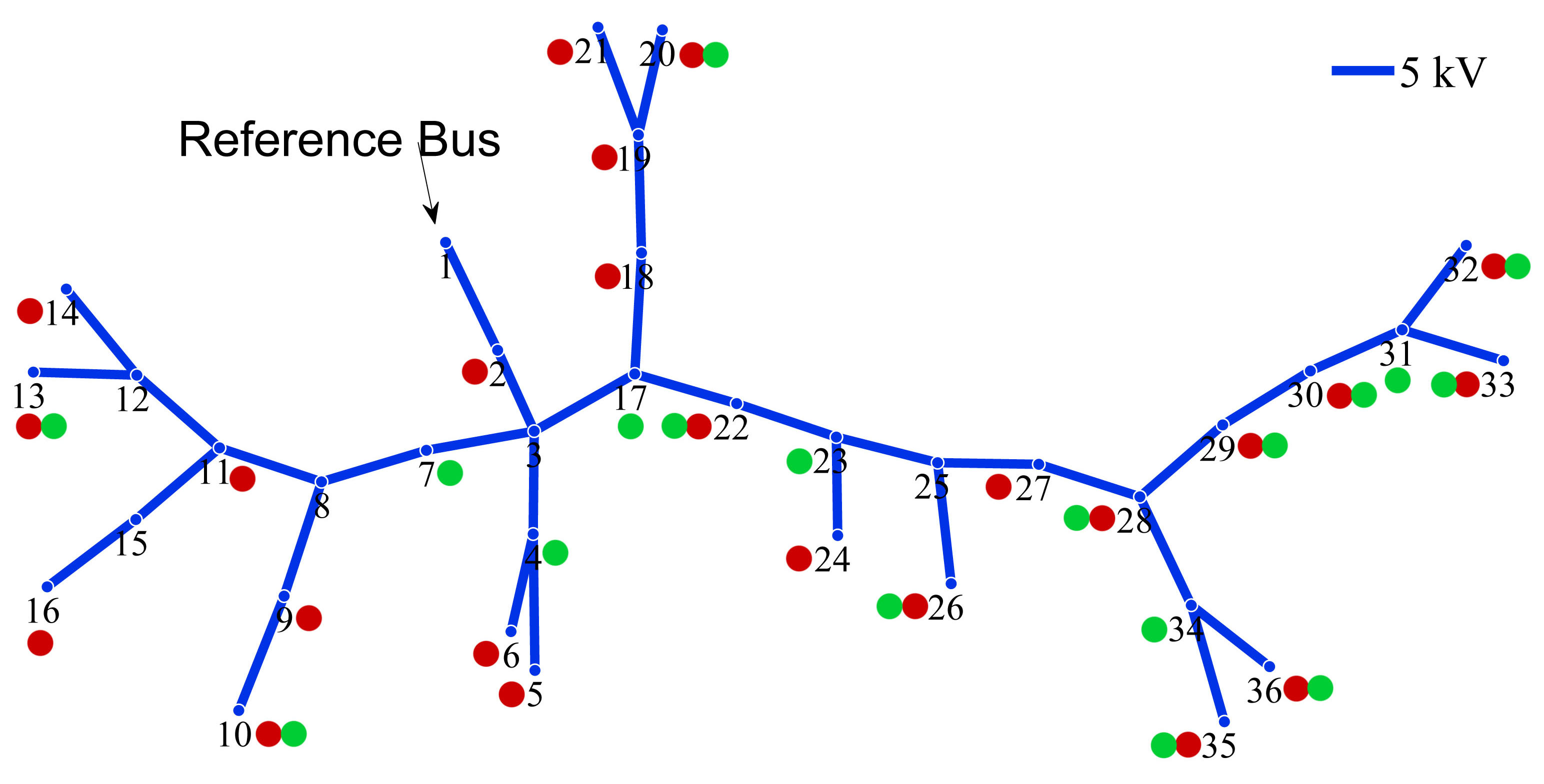}
\caption{The topology of IEEE-37 Node test feeder distribution grid, consisting of 1 substation (also referred to as a reference bus) and 35 additional regular buses. The graph edges stand for 5kV branches. The reference bus is marked, as well as all the nodes that were assigned as loads (in red) and as generators (in green).} \label{fig:ieee37topology}
\end{figure}

\begin{figure}[!ht]
    \centering
    \Mysubfloat{Re\{Y\}}{0.245}{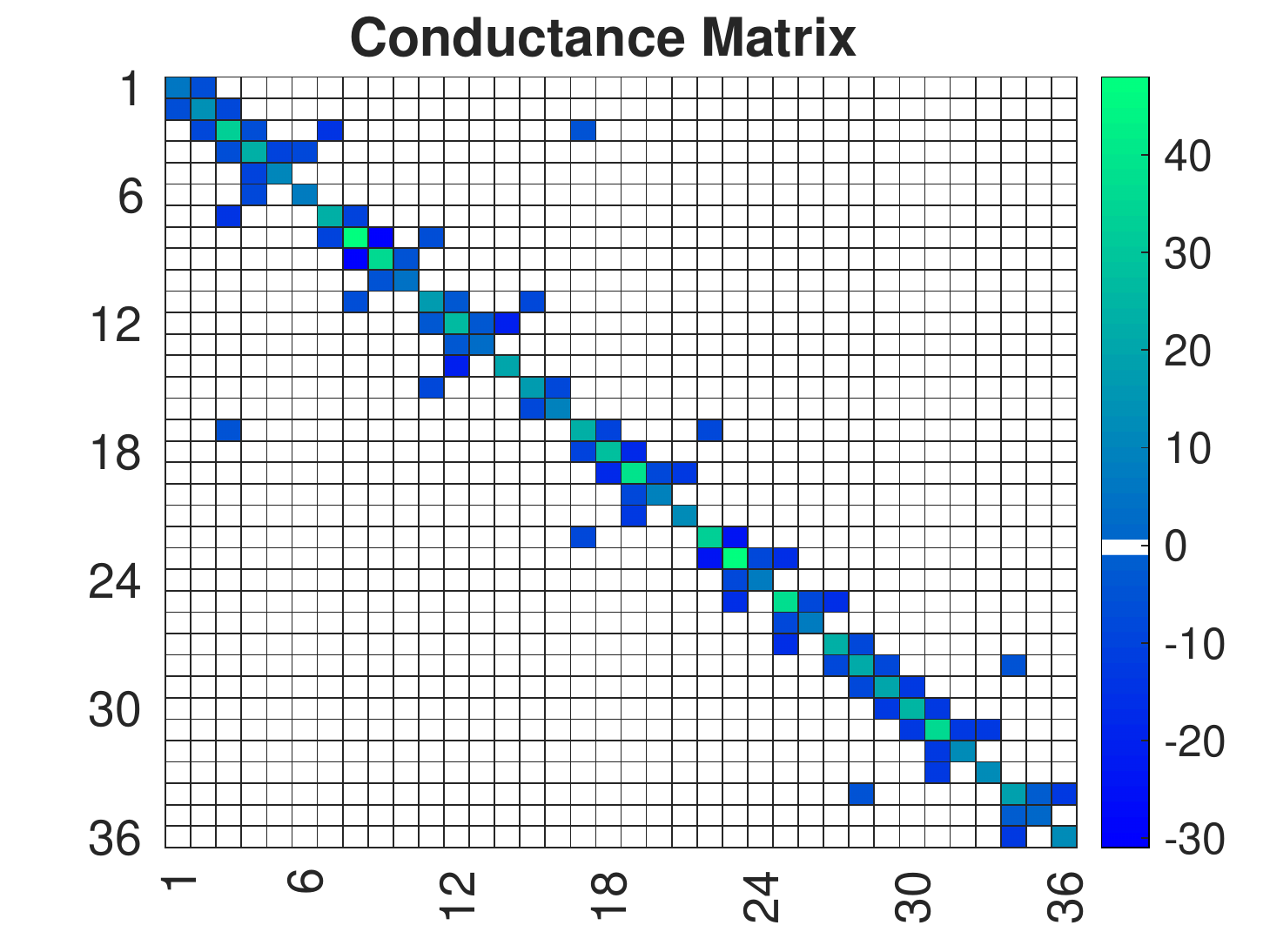}{fig:Y37real}
    \Mysubfloat{Im\{Y\}}{0.245}{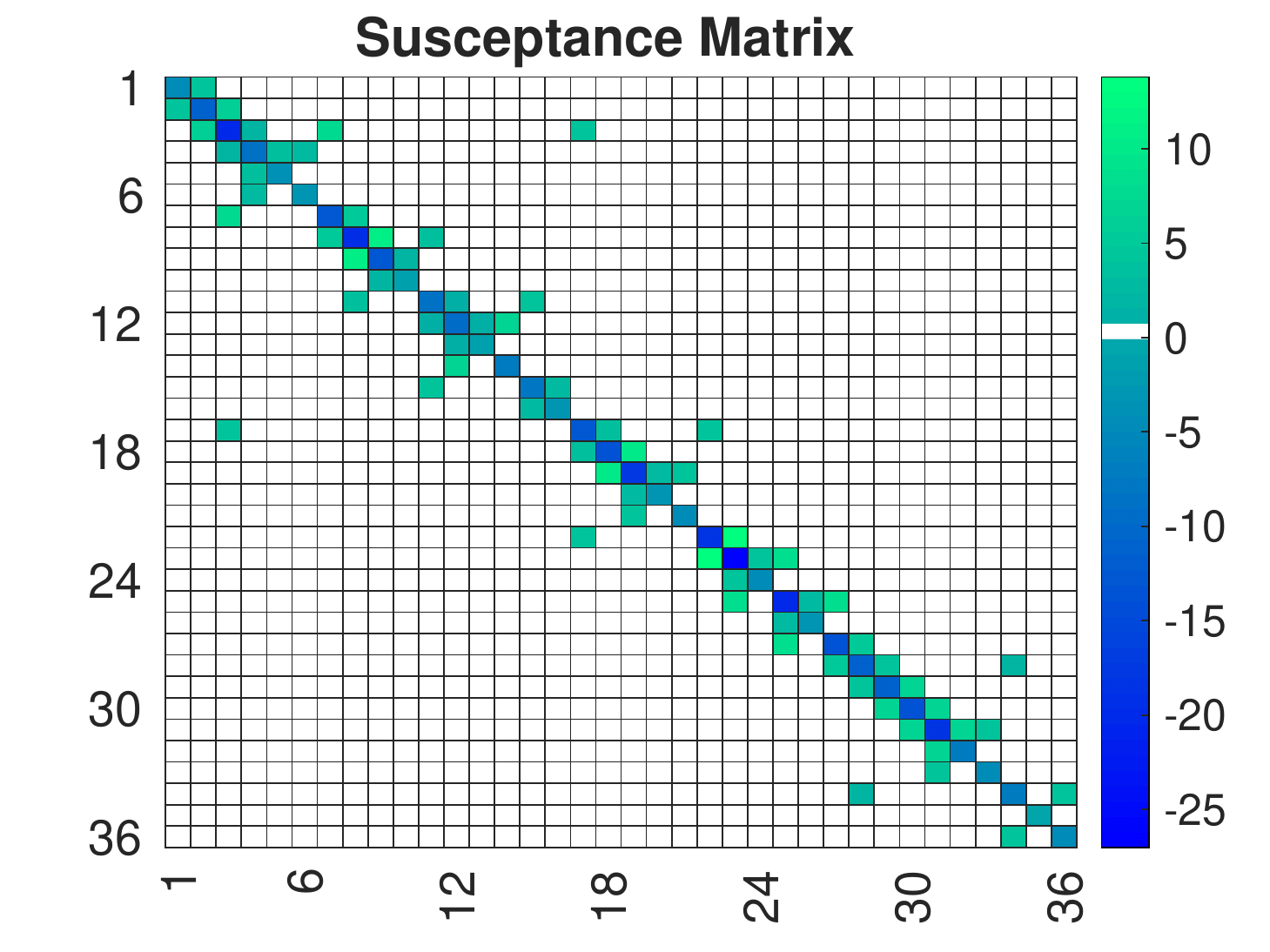}{fig:Y37imag}
    \caption{The Admittance Matrix, $Y$, for the IEEE 37-Node test feeder distribution grid, divided into the real part (a) , and the imaginary part (b) (both measured in Mega-Siemens).}
    \label{fig:Y37}
\end{figure}

Using these measurements, we implemented a test-case scenario, based on a
modified single-phase IEEE 37-node test feeder as follows:
\begin{enumerate}
    \item{For each of the eight reported active power time series, a random
        power-factor value was generated (between 0.96 and 0.98), and used to
        calculate a corresponding time series of the reactive power
        components, establishing eight complex power phasors,
        $s_i(t);i\in{[1,\ldots,8]}$;}
    \item{25 of the buses of the test case were assigned with one of the
        eight available power-phasors, $s_i(t)$. Thus, the eight
        power-phasors were duplicated (in a circular manner) and incorporated
        randomly to 25 of the buses;}
    \item{Similarly, a power phasor was created from the available PV-panel
        generated\footnote{Note that the generation power phasor is created
        with a negative sign, to indicate generation of power.} power
        measurements, $s_{pv}(t)$, and was duplicated and randomly assigned
        onto 18 buses;}
    \item{In case of buses that were assigned both a load power phasor,
        $s_i(t)$, and the generation power phasor, $s_{pv}(t)$, the two
        phasors were summed.}
\end{enumerate}
An illustration of the test case, including the nodes which were chosen as
loads and as generators (based on available measured time series), is depicted
in {\textbf{Fig. \ref{fig:ieee37topology}}}. The admittance matrix, $Y$, of the
same test case is depicted in {\textbf{Fig. \ref{fig:Y37}}}.

\begin{figure}
    \centering
    \Mysubfloat{Pd buses 1-18}{0.245}{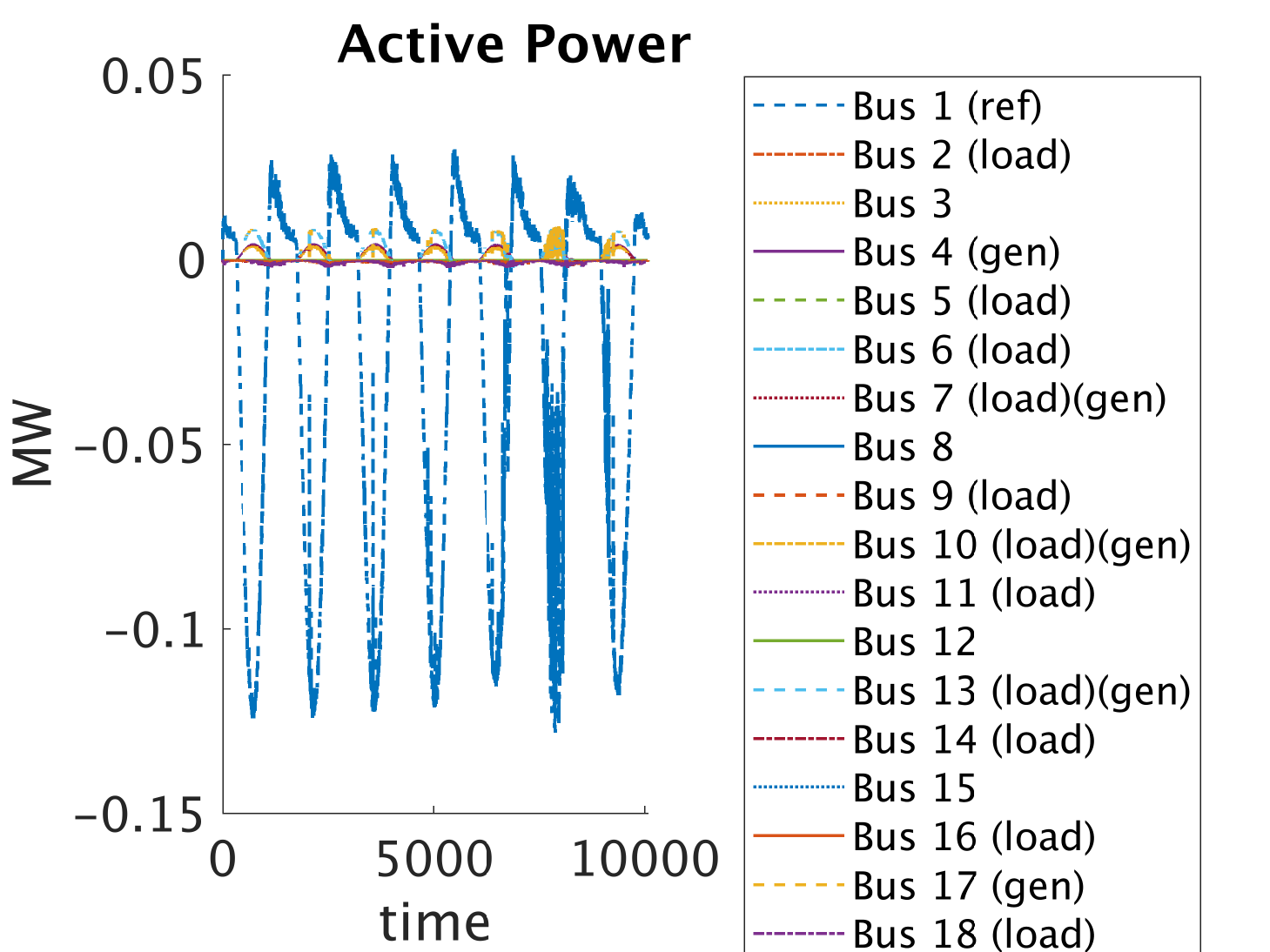}{fig:ieee37_1-18pd_smoothed_downsampled}
    \Mysubfloat{Qd buses 1-18}{0.245}{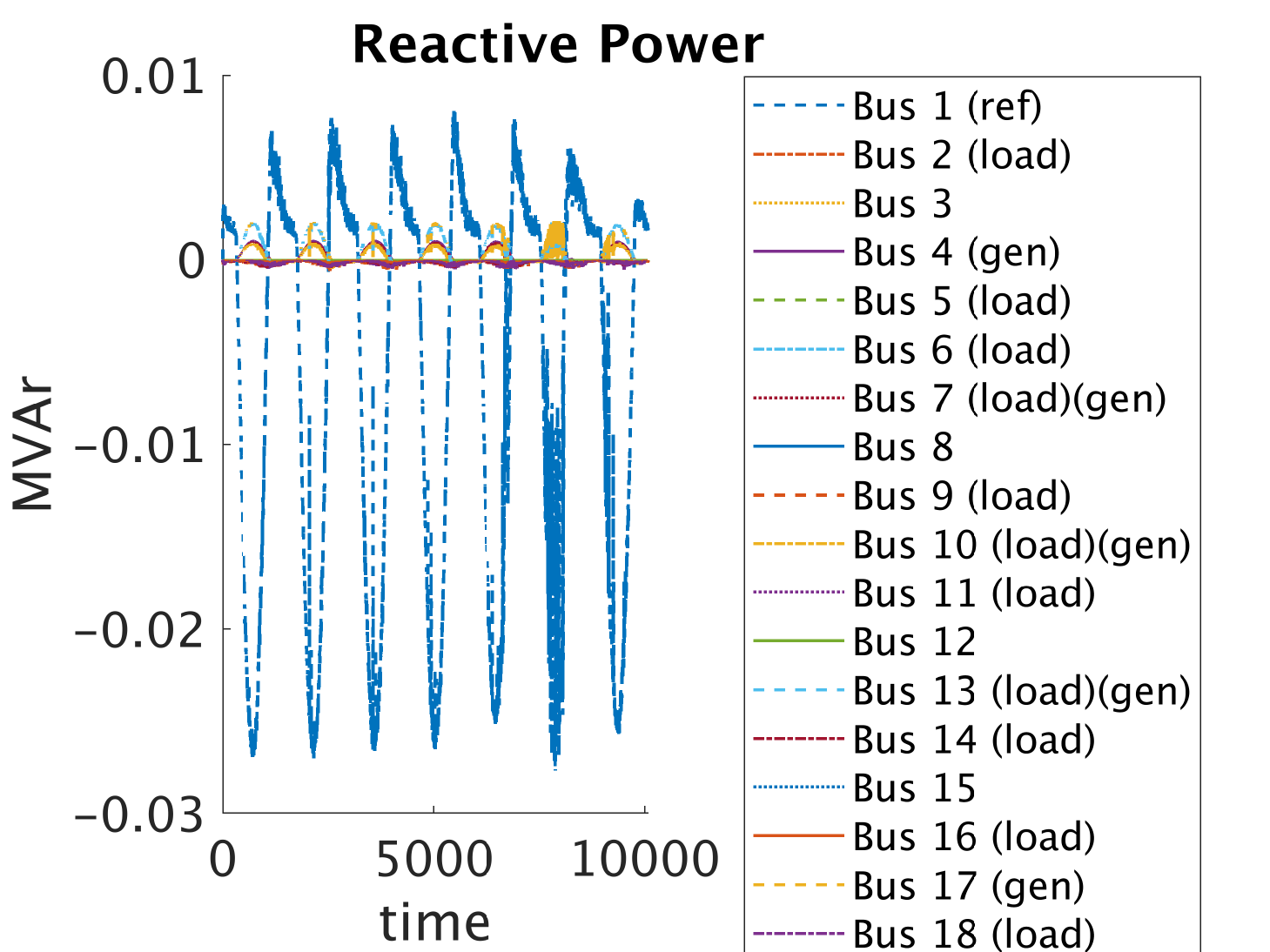}{fig:ieee37_1-18qd_smoothed_downsampled}

    \Mysubfloat{Pd buses 19-36}{0.245}{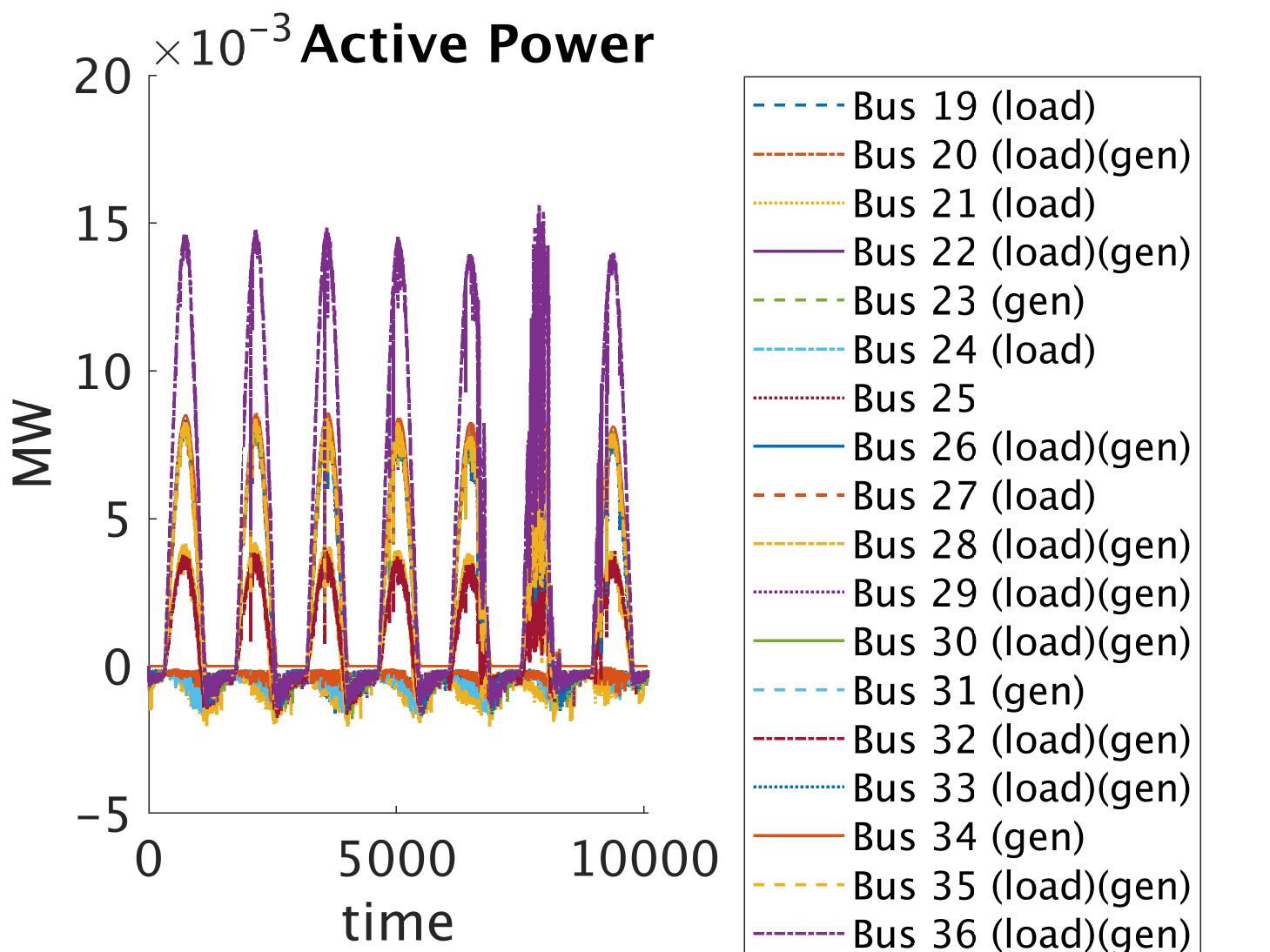}{fig:ieee37_19-36pd_smoothed_downsampled}
    \Mysubfloat{Qd buses 19-36}{0.245}{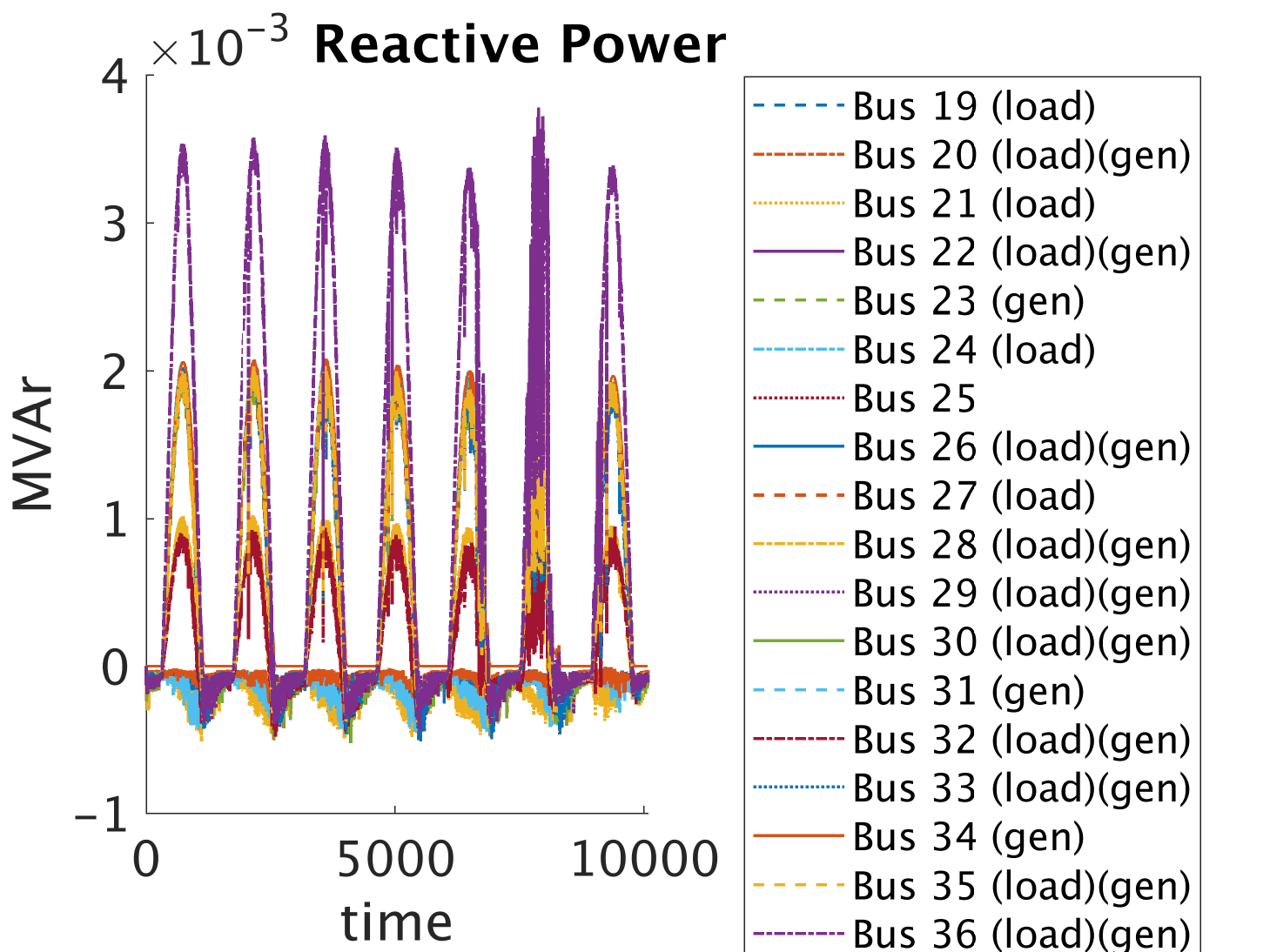}{fig:ieee37_19-36qd_smoothed_downsampled}

    \Mysubfloat{Vm buses 1-18}{0.245}{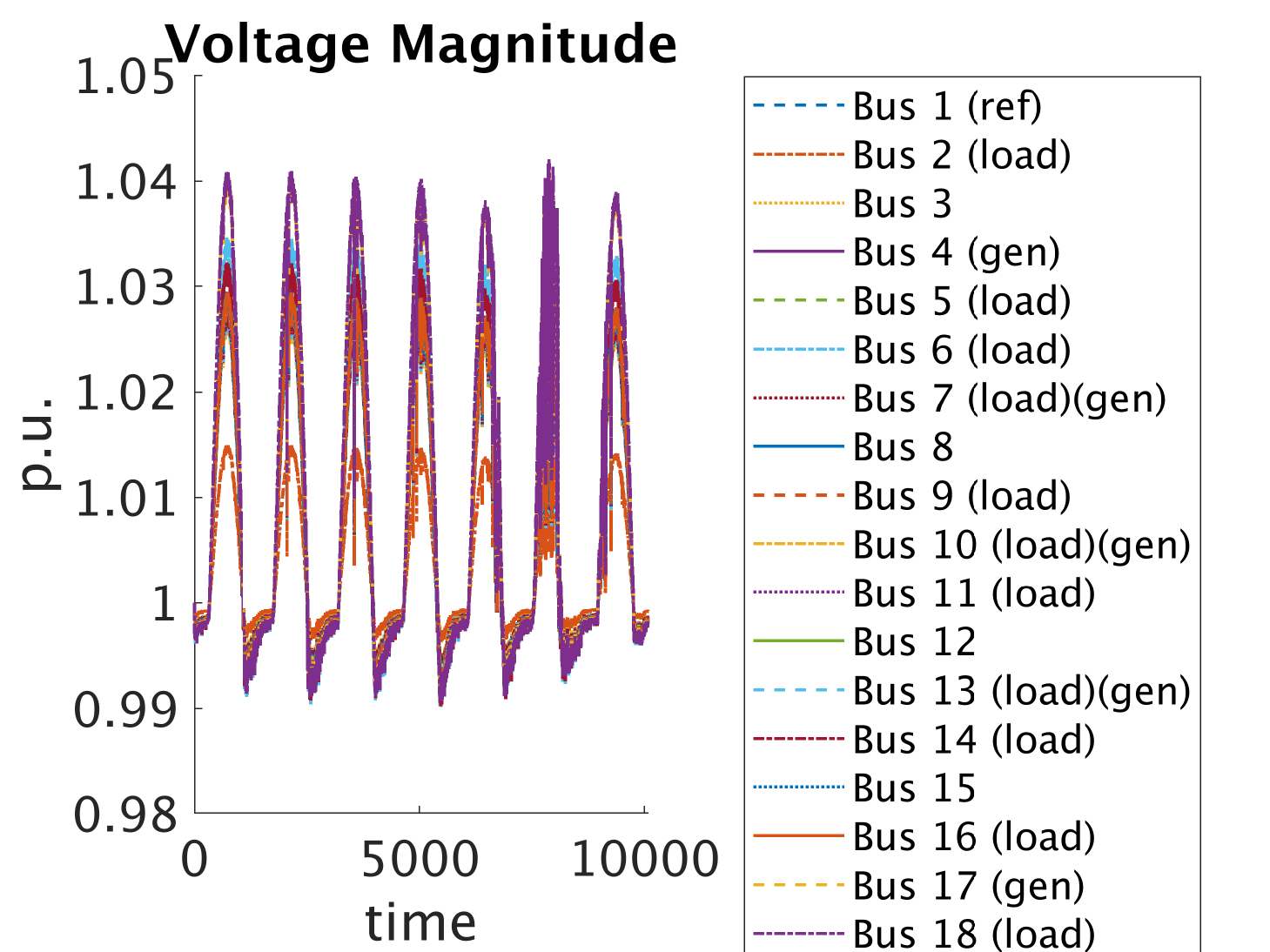}{fig:ieee37_1-18vm_smoothed_downsampled}
    \Mysubfloat{Va buses 1-18}{0.245}{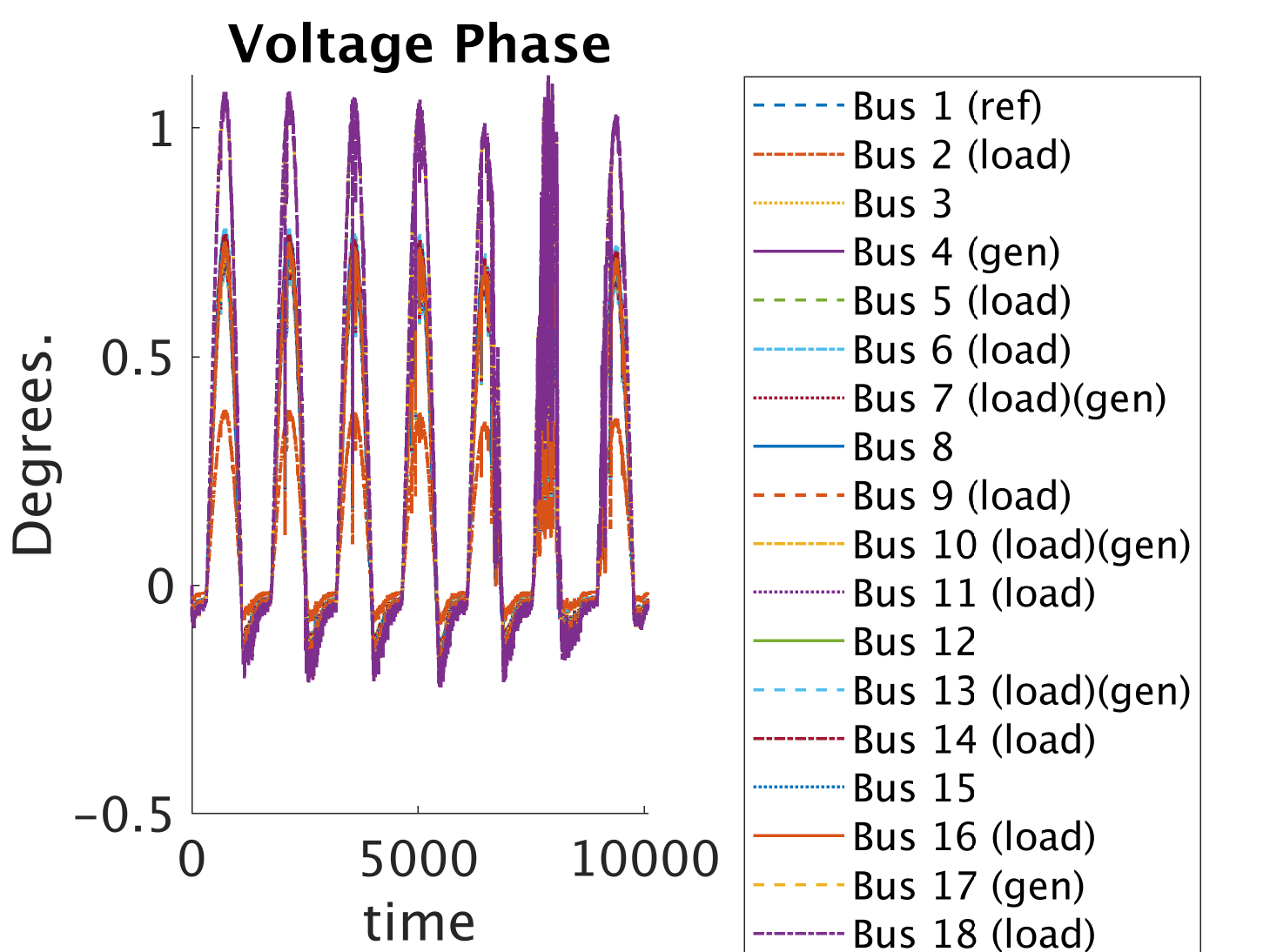}{fig:ieee37_1-18va_smoothed_downsampled}

    \Mysubfloat{Vm buses 19-36}{0.245}{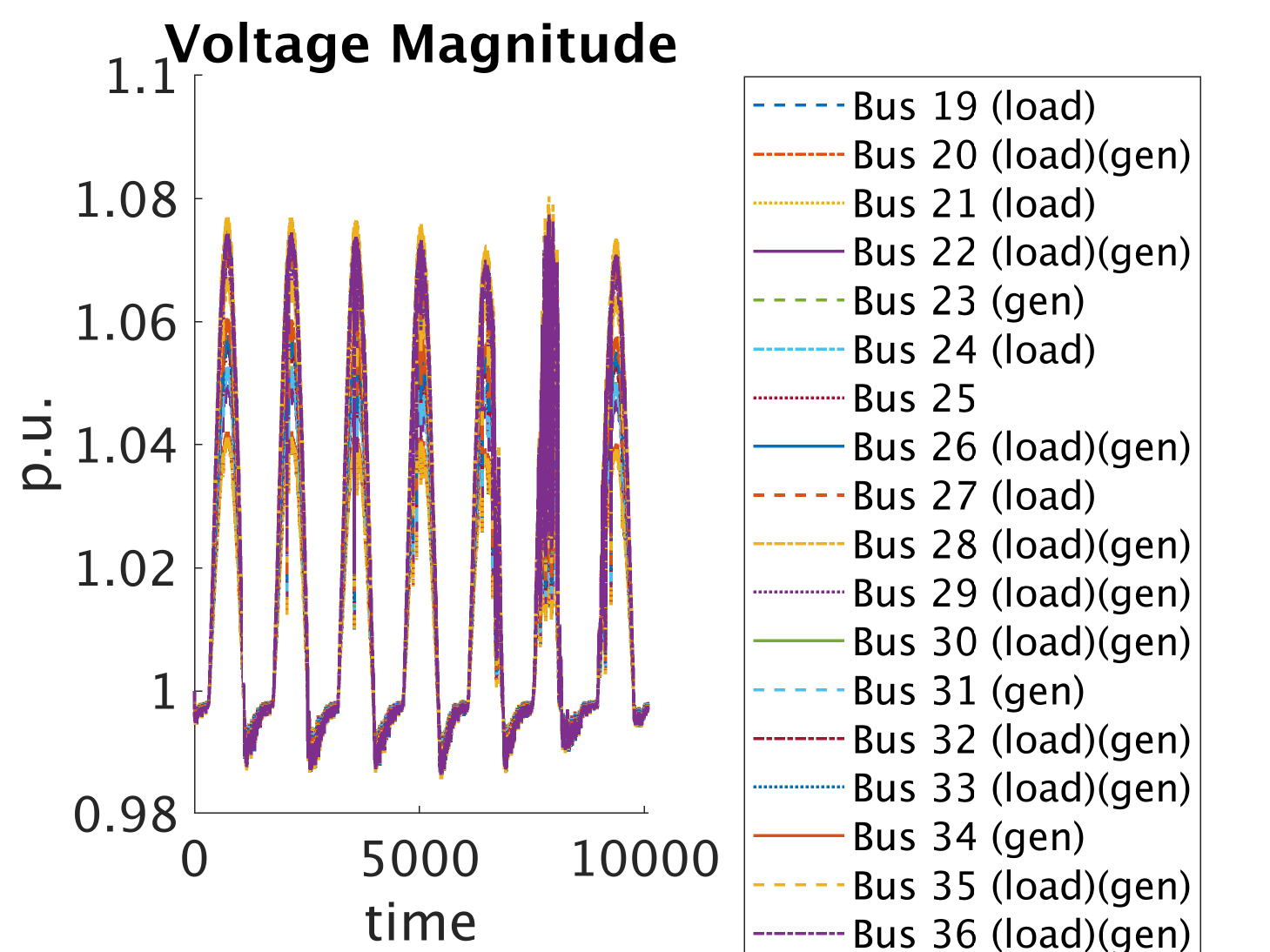}{fig:ieee37_19-36vm_smoothed_downsampled}
    \Mysubfloat{Va buses 19-36}{0.245}{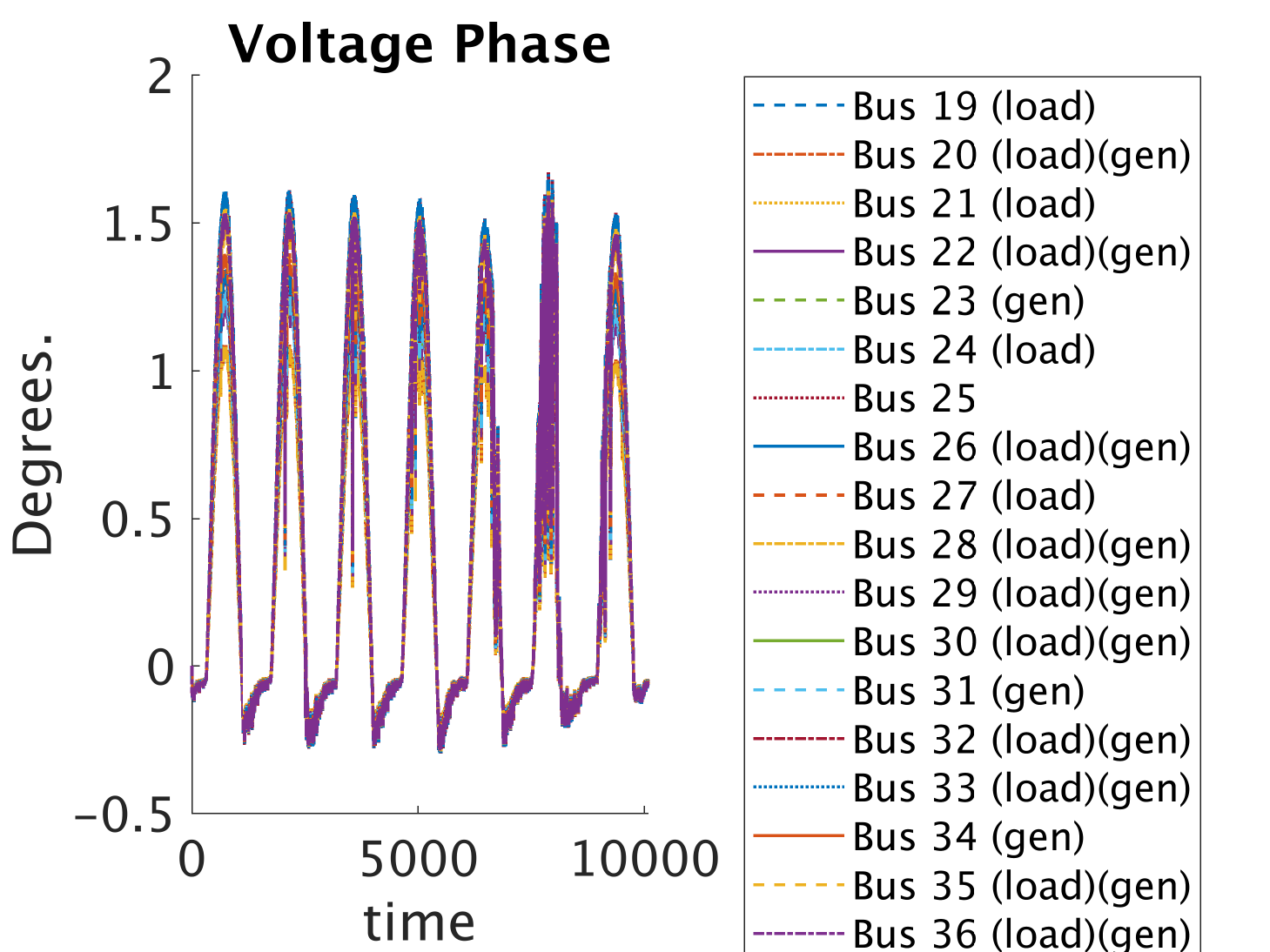}{fig:ieee37_19-36va_smoothed_downsampled}

    \caption{{Smoothed and downsampled} time series of the voltage time series and the power time series as solved by Matpower for the PFE of the IEEE-37 grid. Panels (a),(c) and (b),(d) depict the active and reactive power of buses 1-18 and 19-36, respectively. Panels (e),(g) and (f),(h) depict the voltage magnitudes and angles of buses 1-18 and 19-36, respectively. 1 p.u. equals to $4.8$ KV.}
    \label{fig:data1}
\end{figure}

After acquiring all the time series (i.e., the phasor vectors
$\underline{s}(t)$ with measurements for each of the buses in the test case),
we smoothed the data using a 60-sample moving average window, and then
downsampled the time series by a factor 60. In other words, we smoothed the
data using a low-pass-filter (a moving average over 60 values, sampled at $1$
Hz), and then took one sample per minute. By doing so, we: 1) Reduced the
granularity of the data from 1-second to 1-minute, which reduces the complexity
of the estimation workflows; 2) By smoothing the data prior to the downsampling
operation, we minimized the chance of possible aliasing, and at the same time
dampened the measurement- and other additive-noise by a factor of 60
\cite{oppenheim1999discrete}. As a result, the time series length was reduced
to 10080 time steps. Using the smoothed and downsampled power-phasors time
series, we utilized \emph{Matpower} \cite{zimmerman2010matpower} power-flow
solver to generate the corresponding phasors $\underline{s}(t)$ and
$\underline{v}(t)$ for all the time steps available. The Matpower calculation
was done based on a given and known admittance matrix, $Y$, using the full
AC-model PFE. The Matpower output phasors are plotted in {\textbf{Fig.
\ref{fig:data1}}}.

Based on the Matpower outputs, which include the full pre-processed time series
of power and voltage phasors, we construct a data set for every pair of
$\{T,\numpow(t)\}$, which will be examined in the sequel. Specifically, we
focus on the non observable situation of the SFSE scenario
($\obs^t<\frac{1}{2}$ i.e., $\obs^t<50\%$), where the number of the observable
voltages at the time step $t$ is $N_v(t)=0$, and the number of the observable
power phasors at the same time step is $0\le N_s(t)<N$.

\begin{figure}[!ht]
\centering
\includegraphics[width=90mm]{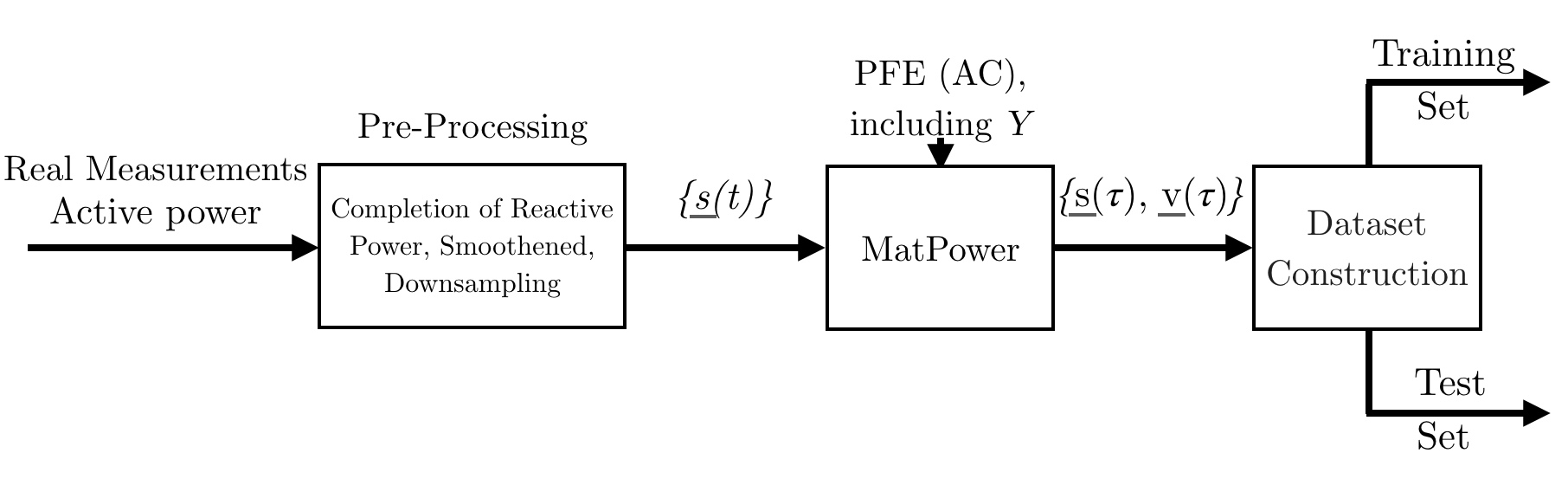}
\caption{Illustration of the Data Preparation Scheme} \label{fig:scheme}
\end{figure}

The data set construction began with randomly selecting a set of $9,000$
$T-$long sequences of time series from the available week-long data set.
$8,100$ sequences (i.e., $90\%$) are drawn from the first six days, and are
used for training the DNN, whereas the other $900$ sequences are drawn from the
last day, to serve as a test set. Each sequence consists of (i) a fully
observable $[T-1]$-long time series containing the power and voltage phasors
for all nodes (i.e., $\underline{s}(\tau)$ and $\underline{v}(\tau)$ for
$\tau{}\in{\{t-1-T,\cdots,t-1\}}$), (ii) a partially observable time step
containing $\numpow(t)$ power phasors, $\underline{s}(t)$, and (iii) a
corresponding target voltage vector, $\underline{v}(t)$. Each power and voltage
component (real and imaginary) in the data set was standardized using the mean
and the standard deviation obtained from the training set\footnote{The
standardization process conducted created the data sets to have a mean of $0$
and a variance of $1.0$.}. An illustration of the full data-preparation scheme
is depicted in {\textbf{Fig. \ref{fig:scheme}}}.

\subsection{Training} 
\begin{figure}[!ht]
\centering
\includegraphics[width=90mm]{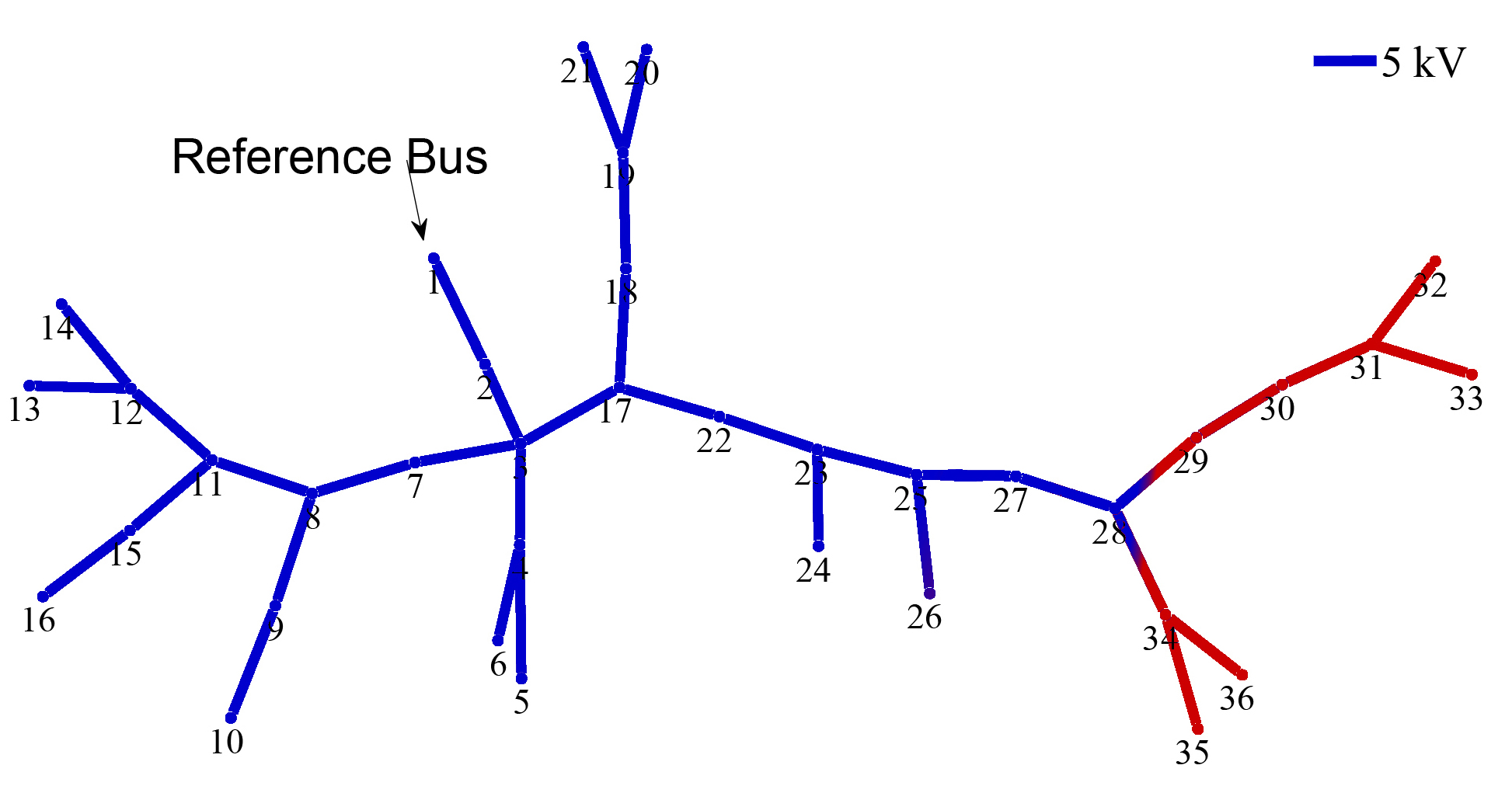}
\caption{Illustration of the IEEE-37 Node test feeder, where none of the
voltages phasors, $\underline{v}(t)$, are known, and part of the power
information, $\underline{s}(t)$ is missing (the buses which do not report the
power information are colored in red), giving an observability value of
$\obs^t=39\%$ (i.e., $\numpow(t)=28; \numvol(t)=0$). The buses that do not
report the power information are selected based on their identification number
(1-36).} \label{fig:ieee37topologySecond}
\end{figure}

We trained DNN using permutations of the following design parameters:
\begin{enumerate}
    \item $T\in\{5,50\}$
    \item $\numpow(t)\in\{35,28,18,12,6\}$
    \item $\lambda\in\{0,1,2,20\}$
\end{enumerate}
giving a total of $40$ different permutations which cover $5$ different
observability scenarios: $\obs(35,0)=49\%$, $\obs(28,0)=39\%$,
$\obs(18,0)=25\%$, $\obs(12,0)=17\%$, $\obs(6,0)=8\%$. In order to simulate a
real-world scenario, increasing the number of buses that do not report the
measurements (and thus, reducing the observability) was not selected randomly,
but rather the observability was lost in a systematic method, based on the
topological formation of the grid. That is, decreasing the observability was
conducted by removing the information from buses located in the same area of
the grid, starting from bus $\#36$ and advancing towards the reference bus,
$\#1$. An example of $\numpow(t)=28$ is illustrated in {\textbf{Fig.
\ref{fig:ieee37topologySecond}}}.

The training of the DNN for each of the different permutations was performed
$30$ times, in order to obtain statistically significant results (each run with
randomly selected (and thus, different) DNN weights initializations). The
optimizer that was used during the training is the Adam
optimizer~\cite{kingma2014adam}, based on mini-batches of $50$ examples.

It is worth noting, that prior to the establishment of the full experimental
setup, we conducted a trace-based simulation (based on parts of the available
processed data sets) using \emph{Matpower} \texttt{case4\underline{~}dist}
4-buses distribution network scenario. Based on this simple trace-based
simulation, we found the general values of the model design parameters (i.e.,
Number of layers in the fully connected modules, feature vector sizes,
activation functions, $\lambda$, $T$), which was later used to design the full
experiment. In other words, we preserved the same DNN architecture and design
parameters of the 4-buses distribution network scenario when implementing the
full experimental setup, up to minor feature vector resizing (due to the
different input size).

\subsection{Evaluation}
We evaluated the trained DNNs using the test-set data (which was taken from the
last $1/7$ part ($900$ sequences) of the real-world data, and was not used for
training). We then calculated the MSE of the resulting voltages estimates (for
each of the permutations, the MSE was obtained by averaging across 30 runs).
Although the data set is standardized and its values are in the rectangular
complex format, we converted the MSE with respect to a de-standardized, polar
representation (Magnitude, Angle)
since this representation is more meaningful for practical uses. %Namely,
the estimated voltages are first de-standardized (by multiplying them by the
standard deviation and adding a mean value) and then transformed to a polar
format. Moreover, since we observed that the typical MSE of the magnitude
values of the estimated voltages differs from the typical MSE of the angle of
the estimated voltages by several orders of magnitudes, we analyzed the
magnitude and the angle separately.

\subsection{Results}
\begin{figure}[!ht]
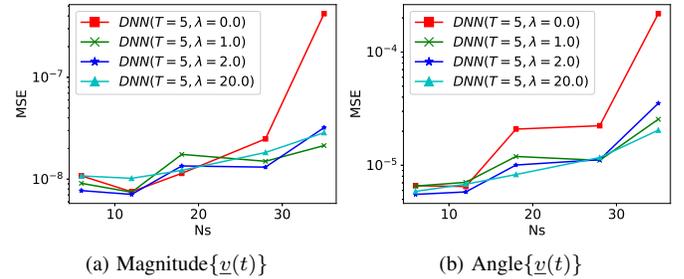

    \centering
    \Mysubfloat{Magnitude\{$\underline{v}(t)$\}}{0.245}{Compare_T5_NN_lambdas_Power_Observability_Sweep_bs50_Nv0_mse_mag}{fig:Compare_T5_NN_lambdas_mag}
    \Mysubfloat{Angle\{$\underline{v}(t)$\}}{0.245}{Compare_T5_NN_lambdas_Power_Observability_Sweep_bs50_Nv0_mse_ang}{fig:Compare_T5_NN_lambdas_ang}
    \caption{{MSE} for the IEEE-37 Node test feeder voltages magnitude and angle estimation under partial observability. This plot compares DNN models trained with different degree of PFE regularization.}
    \label{fig:Compare_T5_NN_lambdas}
\end{figure}

\begin{figure}[!ht]
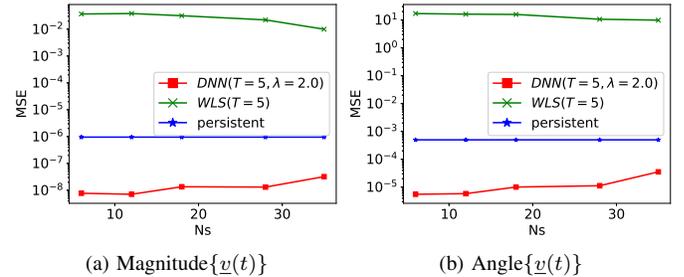

    \centering
    \Mysubfloat{Magnitude\{$\underline{v}(t)$\}}{0.245}{Compare_T5_NN_WLS_Persistent_Power_Observability_Sweep_bs50_Nv0_mse_mag}{fig:Compare_T5_NN_WLS_Persistent_mag}
    \Mysubfloat{Angle\{$\underline{v}(t)$\}}{0.245}{Compare_T5_NN_WLS_Persistent_Power_Observability_Sweep_bs50_Nv0_mse_ang}{fig:Compare_T5_NN_WLS_Persistent_ang}
    \caption{{MSE} for the IEEE-37 Node test feeder voltages magnitude and angle estimation under partial observability - WLS estimation and persistent guess. The DNN results (for $\lambda=2$) are plotted for comparison.}
    \label{fig:Compare_T5_NN_WLS_Persistent}
\end{figure}

The MSE values of the different estimates $\underline{\hat{v}}(t)$ for all
permutations (for $T=5$) are plotted in {\textbf{Fig.
\ref{fig:Compare_T5_NN_lambdas}}}. The MSE values as achieved by the standard
WLS, including a persistence guess for all permutations (for $T=5$) are plotted
in {\textbf{Fig. \ref{fig:Compare_T5_NN_WLS_Persistent}}}.

Looking at {\textbf{Fig. \ref{fig:Compare_T5_NN_lambdas}}} and {\textbf{Fig.
\ref{fig:Compare_T5_NN_WLS_Persistent}}}, we can clearly see that the DNN
outperforms both the WLS-based estimation and the persistence guess by a big
margin. Furthermore, the incorporation of the PFE information into the loss
function as a regularizer (see \eqref{eqloss}) further improves the accuracy of
the estimates. This is  especially evident in the estimation of the
voltage-angles.

\subsubsection{Impact of the selected value of $\lambda$}
A DNN trained with PFE regularization ($\lambda>0$) showed, in general, lower
MSE when compared with a non-regularized DNN ($\lambda=0$).  This phenomenon is
especially evident in the angle estimation (see {\textbf{Fig.
\ref{fig:Compare_T5_NN_WLS_Persistent}}}). Indeed, the improvement is less
pronounced for the magnitude estimation. However, this can be explained by the
fact that the MSE achieved by all DNN's (including $\lambda=0$) is extremely
small, and thus, the overall margin of improvement is narrower.

\subsubsection{Influence of the size of $T$}
\begin{figure}[!ht]
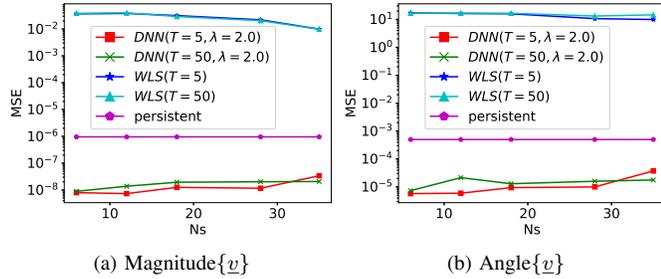

    \centering
    \Mysubfloat{Magnitude\{$\underline{v}$\}}{0.245}{Compare_T5_T50_Power_Observability_Sweep_bs50_Nv0_mse_mag}{fig:Compare_T5_T50_mag}
    \Mysubfloat{Angle\{$\underline{v}$\}}{0.245}{Compare_T5_T50_Power_Observability_Sweep_bs50_Nv0_mse_ang}{fig:Compare_T5_T50_ang}
    \caption{{Impact of $T$}. The Y axis shows the MSE for the IEEE-37 Node
    test feeder voltages magnitude and angle estimation under partial
    observability. The plot compares the estimation MSE of DNN to WLS and
    persistent estimators. All the estimators were checked both with
    $T\in\{5,50\}$.} \label{fig:Compare_T5_T50}
\end{figure}

{\textbf{Fig. \ref{fig:Compare_T5_T50}}} presents a comparison between $T=5$
and $T=50$ for selected permutations. As can be seen, the differences are
negligible. Thus, it can be concluded that the amount of information from
historic data (with respect to future voltage-phasor estimation) is negligible
beyond at least a 5-minute window ($T=5$).

\section{Conclusion and Discussion}
In this paper, we presented a new approach of state-estimation in the
distribution grids during sudden failures or attacks. The method capitalizes on
a physics-informed DNN training algorithm that is able to take advantage of the
grid physical information. We demonstrated the performance of the proposed
method using an experimental setup which simulates a case of a sudden failure
and loss of observability. We showed that our  DNN-based estimation achieves a
higher accuracy of the voltage-phasors estimation when compared with the widely
used WLS-based estimation. Furthermore, the main contribution of incorporating
the PFE regularization into the DNN model was shown to be in the voltage-angles
estimation. The latter is typically overlooked in standard DSSE algorithms, but
will become an important factor in modern and future low-inertia distribution
grids.

Some ideas for further research follow: 1) Although our PFE-induced training of
a DNN showed higher state estimation accuracies, we did not perform a full
analysis study regarding the optimal value of $\lambda$ parameter for a given
observability value ($\obs^t$). The next step of our work is to establish a set
of optimal $\lambda$ values. 2) Furthermore, in this paper we compared our
physics-informed DNN output estimation accuracy to the standard WLS methodology
which uses historic data in order to establish the different weighting of the
missing samples. However, new approaches for the missing PMU data recovery
\cite{Liao2019} may be used in order to enhance both the WLS and our proposed
DNN, which could improve the overall estimation accuracy, and should be
investigated in future work. 3) In this research, we used the DNN model with
real numbers. As the complex-number capable DNNs are currently being
studied~\cite{trabelsi2017deep}, it is worth investigating the DSSE problem
based on a DNN over the $\mathbb{C}$ field. 4) Lastly, it is important to
develop a unified DNN model that will be capable of dealing with multiple
levels of observability without requiring dedicated training sessions.

\bibliographystyle{IEEEtran}

\bibliography{mybibMar2019Drive,LearnPMU,references}

\end{document}

%% file: HighLevel.tex
% Graphic for TeX using PGF
% Title: C:\Users\Kostya\Dropbox\PhD Research\Power\Figures\HighLevel.dia
% Creator: Dia v0.97.2
% CreationDate: Wed Oct 02 21:51:53 2019
% For: Kostya
% \usepackage{tikz}
% The following commands are not supported in PSTricks at present
% We define them conditionally, so when they are implemented,
% this pgf file will use them.
\ifx\du\undefined
  \newlength{\du}
\fi
\setlength{\du}{15\unitlength}
\begin{tikzpicture}[scale=0.59, every node/.style={scale=0.7}]
\pgftransformxscale{1.000000}
\pgftransformyscale{-1.000000}
\definecolor{dialinecolor}{rgb}{0.000000, 0.000000, 0.000000}
\pgfsetstrokecolor{dialinecolor}
\definecolor{dialinecolor}{rgb}{1.000000, 1.000000, 1.000000}
\pgfsetfillcolor{dialinecolor}
\definecolor{dialinecolor}{rgb}{0.964706, 0.878431, 0.760784}
\pgfsetfillcolor{dialinecolor}
\fill (17.600000\du,16.200000\du)--(17.600000\du,20.600000\du)--(23.362500\du,20.600000\du)--(23.362500\du,16.200000\du)--cycle;
\pgfsetlinewidth{0.100000\du}
\pgfsetdash{}{0pt}
\pgfsetdash{}{0pt}
\pgfsetmiterjoin
\definecolor{dialinecolor}{rgb}{0.000000, 0.000000, 0.000000}
\pgfsetstrokecolor{dialinecolor}
\draw (17.600000\du,16.200000\du)--(17.600000\du,20.600000\du)--(23.362500\du,20.600000\du)--(23.362500\du,16.200000\du)--cycle;
% setfont left to latex
\definecolor{dialinecolor}{rgb}{0.000000, 0.000000, 0.000000}
\pgfsetstrokecolor{dialinecolor}
\node at (20.481250\du,17.440000\du){Fully Observable};
% setfont left to latex
\definecolor{dialinecolor}{rgb}{0.000000, 0.000000, 0.000000}
\pgfsetstrokecolor{dialinecolor}
\node at (20.481250\du,18.240000\du){Measurements};
% setfont left to latex
\definecolor{dialinecolor}{rgb}{0.000000, 0.000000, 0.000000}
\pgfsetstrokecolor{dialinecolor}
\node at (20.481250\du,19.040000\du){};
% setfont left to latex
\definecolor{dialinecolor}{rgb}{0.000000, 0.000000, 0.000000}
\pgfsetstrokecolor{dialinecolor}
\node at (20.481250\du,19.840000\du){\emph{(t-1-T,...,t-1)}};
\pgfsetlinewidth{0.100000\du}
\pgfsetdash{}{0pt}
\pgfsetdash{}{0pt}
\pgfsetbuttcap
\pgfsetmiterjoin
\pgfsetlinewidth{0.100000\du}
\pgfsetbuttcap
\pgfsetmiterjoin
\pgfsetdash{}{0pt}
\definecolor{dialinecolor}{rgb}{1.000000, 0.5, 0.5}
\pgfsetfillcolor{dialinecolor}
\pgfpathmoveto{\pgfpoint{24.400000\du}{16.200000\du}}
\pgfpathlineto{\pgfpoint{26.000000\du}{16.200000\du}}
\pgfpathlineto{\pgfpoint{26.000000\du}{17.914286\du}}
\pgfpathcurveto{\pgfpoint{25.680000\du}{17.628571\du}}{\pgfpoint{25.520000\du}{17.628571\du}}{\pgfpoint{25.200000\du}{17.914286\du}}
\pgfpathcurveto{\pgfpoint{24.880000\du}{18.200000\du}}{\pgfpoint{24.720000\du}{18.200000\du}}{\pgfpoint{24.400000\du}{17.914286\du}}
\pgfpathlineto{\pgfpoint{24.400000\du}{16.200000\du}}
\pgfusepath{fill}
\definecolor{dialinecolor}{rgb}{0.000000, 0.000000, 0.000000}
\pgfsetstrokecolor{dialinecolor}
\pgfpathmoveto{\pgfpoint{24.400000\du}{16.200000\du}}
\pgfpathlineto{\pgfpoint{26.000000\du}{16.200000\du}}
\pgfpathlineto{\pgfpoint{26.000000\du}{17.914286\du}}
\pgfpathcurveto{\pgfpoint{25.680000\du}{17.628571\du}}{\pgfpoint{25.520000\du}{17.628571\du}}{\pgfpoint{25.200000\du}{17.914286\du}}
\pgfpathcurveto{\pgfpoint{24.880000\du}{18.200000\du}}{\pgfpoint{24.720000\du}{18.200000\du}}{\pgfpoint{24.400000\du}{17.914286\du}}
\pgfpathlineto{\pgfpoint{24.400000\du}{16.200000\du}}
\pgfusepath{stroke}
% setfont left to latex
\definecolor{dialinecolor}{rgb}{0.000000, 0.000000, 0.000000}
\pgfsetstrokecolor{dialinecolor}
\node at (25.200000\du,17.154286\du){};
\pgfsetlinewidth{0.100000\du}
\pgfsetdash{}{0pt}
\pgfsetdash{}{0pt}
\pgfsetbuttcap
\pgfsetmiterjoin
\pgfsetlinewidth{0.100000\du}
\pgfsetbuttcap
\pgfsetmiterjoin
\pgfsetdash{}{0pt}
\definecolor{dialinecolor}{rgb}{0.964706, 0.878431, 0.760784}
\pgfsetfillcolor{dialinecolor}
\pgfpathmoveto{\pgfpoint{26.000000\du}{20.600000\du}}
\pgfpathlineto{\pgfpoint{24.400000\du}{20.600000\du}}
\pgfpathlineto{\pgfpoint{24.400000\du}{18.371429\du}}
\pgfpathcurveto{\pgfpoint{24.720000\du}{18.742857\du}}{\pgfpoint{24.880000\du}{18.742857\du}}{\pgfpoint{25.200000\du}{18.371429\du}}
\pgfpathcurveto{\pgfpoint{25.520000\du}{18.000000\du}}{\pgfpoint{25.680000\du}{18.000000\du}}{\pgfpoint{26.000000\du}{18.371429\du}}
\pgfpathlineto{\pgfpoint{26.000000\du}{20.600000\du}}
\pgfusepath{fill}
\definecolor{dialinecolor}{rgb}{0.000000, 0.000000, 0.000000}
\pgfsetstrokecolor{dialinecolor}
\pgfpathmoveto{\pgfpoint{26.000000\du}{20.600000\du}}
\pgfpathlineto{\pgfpoint{24.400000\du}{20.600000\du}}
\pgfpathlineto{\pgfpoint{24.400000\du}{18.371429\du}}
\pgfpathcurveto{\pgfpoint{24.720000\du}{18.742857\du}}{\pgfpoint{24.880000\du}{18.742857\du}}{\pgfpoint{25.200000\du}{18.371429\du}}
\pgfpathcurveto{\pgfpoint{25.520000\du}{18.000000\du}}{\pgfpoint{25.680000\du}{18.000000\du}}{\pgfpoint{26.000000\du}{18.371429\du}}
\pgfpathlineto{\pgfpoint{26.000000\du}{20.600000\du}}
\pgfusepath{stroke}
% setfont left to latex
\definecolor{dialinecolor}{rgb}{0.000000, 0.000000, 0.000000}
\pgfsetstrokecolor{dialinecolor}
\node at (25.200000\du,19.211429\du){PO};
\definecolor{dialinecolor}{rgb}{0.000000, 0.000000, 0.000000}
\pgfsetstrokecolor{dialinecolor}
\node at (25.200000\du,17.011429\du){GT};
% setfont left to latex
\definecolor{dialinecolor}{rgb}{0.000000, 0.000000, 0.000000}
\pgfsetstrokecolor{dialinecolor}
\node at (25.200000\du,20.111429\du){\emph{(t)}};
\definecolor{dialinecolor}{rgb}{0.964706, 0.878431, 0.760784}
\pgfsetfillcolor{dialinecolor}
\fill (27.000000\du,16.200000\du)--(27.000000\du,20.600000\du)--(35.500000\du,20.600000\du)--(35.500000\du,16.200000\du)--cycle;
\pgfsetlinewidth{0.100000\du}
\pgfsetdash{}{0pt}
\pgfsetdash{}{0pt}
\pgfsetmiterjoin
\definecolor{dialinecolor}{rgb}{0.000000, 0.000000, 0.000000}
\pgfsetstrokecolor{dialinecolor}
\draw (27.000000\du,16.200000\du)--(27.000000\du,20.600000\du)--(35.500000\du,20.600000\du)--(35.500000\du,16.200000\du)--cycle;
% setfont left to latex
\definecolor{dialinecolor}{rgb}{0.000000, 0.000000, 0.000000}
\pgfsetstrokecolor{dialinecolor}
\node at (31.250000\du,18.440000\du){\Large DNN};
%\pgfsetstrokecolor{dialinecolor}
%\node at (31.250000\du,19.040000\du){\Large Neural Network};
\definecolor{dialinecolor}{rgb}{1.000000, 0.5, 0.5}
\pgfsetfillcolor{dialinecolor}
\pgfpathellipse{\pgfpoint{38.936911\du}{16.199235\du}}{\pgfpoint{2.223435\du}{0\du}}{\pgfpoint{0\du}{1.245963\du}}
\pgfusepath{fill}
\pgfsetlinewidth{0.100000\du}
\pgfsetdash{}{0pt}
\pgfsetdash{}{0pt}
\pgfsetmiterjoin
\definecolor{dialinecolor}{rgb}{0.000000, 0.000000, 0.000000}
\pgfsetstrokecolor{dialinecolor}
\pgfpathellipse{\pgfpoint{38.936911\du}{16.199235\du}}{\pgfpoint{2.223435\du}{0\du}}{\pgfpoint{0\du}{1.245963\du}}
\pgfusepath{stroke}
% setfont left to latex
\definecolor{dialinecolor}{rgb}{0.000000, 0.000000, 0.000000}
\pgfsetstrokecolor{dialinecolor}
\node at (38.936911\du,15.839235\du){PFE};
% setfont left to latex
\definecolor{dialinecolor}{rgb}{0.000000, 0.000000, 0.000000}
\pgfsetstrokecolor{dialinecolor}
\node at (38.936911\du,16.639235\du){Loss};
\pgfsetlinewidth{0.100000\du}
\pgfsetdash{}{0pt}
\pgfsetdash{}{0pt}
\pgfsetmiterjoin
\pgfsetbuttcap
{
\definecolor{dialinecolor}{rgb}{0.000000, 0.000000, 0.000000}
\pgfsetfillcolor{dialinecolor}
% was here!!!
\pgfsetarrowsend{latex}
{\pgfsetcornersarced{\pgfpoint{0.000000\du}{0.000000\du}}\definecolor{dialinecolor}{rgb}{0.000000, 0.000000, 0.000000}
\pgfsetstrokecolor{dialinecolor}
\draw (25.200000\du,16.200000\du)--(25.200000\du,13.852955\du)--(38.936911\du,13.852955\du)--(38.936911\du,14.902955\du);
}}
\pgfsetlinewidth{0.100000\du}
\pgfsetdash{}{0pt}
\pgfsetdash{}{0pt}
\pgfsetbuttcap
{
\definecolor{dialinecolor}{rgb}{0.000000, 0.000000, 0.000000}
\pgfsetfillcolor{dialinecolor}
% was here!!!
\pgfsetarrowsend{latex}
\definecolor{dialinecolor}{rgb}{0.000000, 0.000000, 0.000000}
\pgfsetstrokecolor{dialinecolor}
\draw (42.200000\du,16.201972\du)--(41.160345\du,16.199235\du);
}
\pgfsetlinewidth{0.100000\du}
\pgfsetdash{}{0pt}
\pgfsetdash{}{0pt}
\pgfsetbuttcap
{
\definecolor{dialinecolor}{rgb}{0.000000, 0.000000, 0.000000}
\pgfsetfillcolor{dialinecolor}
% was here!!!
\pgfsetarrowsend{latex}
\definecolor{dialinecolor}{rgb}{0.000000, 0.000000, 0.000000}
\pgfsetstrokecolor{dialinecolor}
\draw (35.500000\du,19.500000\du)--(37.936911\du,19.500000\du);
}
\pgfsetlinewidth{0.100000\du}
\pgfsetdash{}{0pt}
\pgfsetdash{}{0pt}
\pgfsetbuttcap
{
\definecolor{dialinecolor}{rgb}{0.000000, 0.000000, 0.000000}
\pgfsetfillcolor{dialinecolor}
% was here!!!
\pgfsetarrowsend{latex}
\definecolor{dialinecolor}{rgb}{0.000000, 0.000000, 0.000000}
\pgfsetstrokecolor{dialinecolor}
\draw (38.936911\du,18.600000\du)--(38.936911\du,17.445198\du);
}
\definecolor{dialinecolor}{rgb}{0.964706, 0.878431, 0.760784}
\pgfsetfillcolor{dialinecolor}
\fill (37.936911\du,18.600000\du)--(37.936911\du,20.400000\du)--(39.936911\du,20.400000\du)--(39.936911\du,18.600000\du)--cycle;
\pgfsetlinewidth{0.100000\du}
\pgfsetdash{}{0pt}
\pgfsetdash{}{0pt}
\pgfsetmiterjoin
\definecolor{dialinecolor}{rgb}{0.000000, 0.000000, 0.000000}
\pgfsetstrokecolor{dialinecolor}
\draw (37.936911\du,18.600000\du)--(37.936911\du,20.400000\du)--(39.936911\du,20.400000\du)--(39.936911\du,18.600000\du)--cycle;
% setfont left to latex
\definecolor{dialinecolor}{rgb}{0.000000, 0.000000, 0.000000}
\pgfsetstrokecolor{dialinecolor}
\node at (38.936911\du,19.040000\du){State};
% setfont left to latex
\definecolor{dialinecolor}{rgb}{0.000000, 0.000000, 0.000000}
\pgfsetstrokecolor{dialinecolor}
\node at (38.936911\du,20.00000\du){$\underline{\hat{v}}(t)$};
\pgfsetlinewidth{0.100000\du}
\pgfsetdash{}{0pt}
\pgfsetdash{}{0pt}
\pgfsetbuttcap
{
\definecolor{dialinecolor}{rgb}{0.000000, 0.000000, 0.000000}
\pgfsetfillcolor{dialinecolor}
% was here!!!
\pgfsetarrowsend{latex}
\definecolor{dialinecolor}{rgb}{0.000000, 0.000000, 0.000000}
\pgfsetstrokecolor{dialinecolor}
\draw (26.000000\du,19.485714\du)--(27.000000\du,19.500000\du);
}
\pgfsetlinewidth{0.100000\du}
\pgfsetdash{}{0pt}
\pgfsetdash{}{0pt}
\pgfsetbuttcap
{
\definecolor{dialinecolor}{rgb}{0.000000, 0.000000, 0.000000}
\pgfsetfillcolor{dialinecolor}
% was here!!!
\pgfsetarrowsend{latex}
\definecolor{dialinecolor}{rgb}{0.000000, 0.000000, 0.000000}
\pgfsetstrokecolor{dialinecolor}
\draw (23.362500\du,19.500000\du)--(24.400000\du,19.485714\du);
}
\pgfsetlinewidth{0.100000\du}
\pgfsetdash{}{0pt}
\pgfsetdash{}{0pt}
\pgfsetbuttcap
{
\definecolor{dialinecolor}{rgb}{0.000000, 0.000000, 0.000000}
\pgfsetfillcolor{dialinecolor}
% was here!!!
\pgfsetarrowsend{latex}
\definecolor{dialinecolor}{rgb}{0.000000, 0.000000, 0.000000}
\pgfsetstrokecolor{dialinecolor}
\draw (36.663817\du,16.199741\du)--(35.500000\du,16.200000\du);
}
% setfont left to latex
\definecolor{dialinecolor}{rgb}{0.000000, 0.000000, 0.000000}
\pgfsetstrokecolor{dialinecolor}
\node[anchor=west] at (31.000000\du,14.500000\du){power};
% setfont left to latex
\definecolor{dialinecolor}{rgb}{0.000000, 0.000000, 0.000000}
\pgfsetstrokecolor{dialinecolor}
\node[anchor=west] at (39.200000\du,18.000000\du){voltage};
\definecolor{dialinecolor}{rgb}{1.000000, 0.5, 0.5}
\pgfsetfillcolor{dialinecolor}
\fill (42.200000\du,14.800000\du)--(42.200000\du,17.500000\du)--(45.980000\du,17.500000\du)--(45.980000\du,14.800000\du)--cycle;
\pgfsetlinewidth{0.100000\du}
\pgfsetdash{}{0pt}
\pgfsetdash{}{0pt}
\pgfsetmiterjoin
\definecolor{dialinecolor}{rgb}{0.000000, 0.000000, 0.000000}
\pgfsetstrokecolor{dialinecolor}
\draw (42.200000\du,14.800000\du)--(42.200000\du,17.500000\du)--(45.980000\du,17.500000\du)--(45.980000\du,14.800000\du)--cycle;
% setfont left to latex
\definecolor{dialinecolor}{rgb}{0.000000, 0.000000, 0.000000}
\pgfsetstrokecolor{dialinecolor}
\node at (44.040000\du,15.390000\du){Admittance};
% setfont left to latex
\definecolor{dialinecolor}{rgb}{0.000000, 0.000000, 0.000000}
\pgfsetstrokecolor{dialinecolor}
\node at (44.040000\du,16.190000\du){Matrix};
% setfont left to latex
\definecolor{dialinecolor}{rgb}{0.000000, 0.000000, 0.000000}
\pgfsetstrokecolor{dialinecolor}
\node at (44.040000\du,17.00000\du){Y};
% setfont left to latex
\definecolor{dialinecolor}{rgb}{0.000000, 0.000000, 0.000000}
\pgfsetstrokecolor{dialinecolor}
\node[anchor=west] at (17.600000\du,14.600000\du){Power/Voltage};
\definecolor{dialinecolor}{rgb}{0.000000, 0.000000, 0.000000}
\pgfsetstrokecolor{dialinecolor}
\node[anchor=west] at (17.600000\du,15.600000\du){Measurements};
\end{tikzpicture}

%% file: NN_Architecture.tex
% Graphic for TeX using PGF
% Title: D:\Dropbox\PhD Research\Power\Figures\NN_Architecture.dia
% Creator: Dia v0.97.2
% CreationDate: Fri Sep 27 11:20:17 2019
% For: Orenb
% \usepackage{tikz}
% The following commands are not supported in PSTricks at present
% We define them conditionally, so when they are implemented,
% this pgf file will use them.
\ifx\du\undefined
  \newlength{\du}
\fi
\setlength{\du}{15\unitlength}
\begin{tikzpicture}[thick,scale=0.4, every node/.style={scale=0.55}]
\pgftransformxscale{1.000000}
\pgftransformyscale{-1.000000}
\definecolor{dialinecolor}{rgb}{0.000000, 0.000000, 0.000000}
\pgfsetstrokecolor{dialinecolor}
\definecolor{dialinecolor}{rgb}{1.000000, 1.000000, 1.000000}
\pgfsetfillcolor{dialinecolor}
\definecolor{dialinecolor}{rgb}{0.964706, 0.878431, 0.760784}
\pgfsetfillcolor{dialinecolor}
\fill (35.500000\du,12.000000\du)--(35.500000\du,28.000000\du)--(50.000000\du,28.000000\du)--(50.000000\du,12.000000\du)--cycle;
\pgfsetlinewidth{0.100000\du}
\pgfsetdash{}{0pt}
\pgfsetdash{}{0pt}
\pgfsetmiterjoin
\definecolor{dialinecolor}{rgb}{0.000000, 0.000000, 0.000000}
\pgfsetstrokecolor{dialinecolor}
\draw (35.500000\du,12.000000\du)--(35.500000\du,28.000000\du)--(50.000000\du,28.000000\du)--(50.000000\du,12.000000\du)--cycle;
% setfont left to latex
\definecolor{dialinecolor}{rgb}{0.000000, 0.000000, 0.000000}
\pgfsetstrokecolor{dialinecolor}
\node at (42.750000\du,20.528056\du){};
\definecolor{dialinecolor}{rgb}{0.921569, 0.945098, 0.874510}
\pgfsetfillcolor{dialinecolor}
\fill (29.000000\du,14.000000\du)--(29.000000\du,17.500000\du)--(35.000000\du,17.500000\du)--(35.000000\du,14.000000\du)--cycle;
\pgfsetlinewidth{0.100000\du}
\pgfsetdash{}{0pt}
\pgfsetdash{}{0pt}
\pgfsetmiterjoin
\definecolor{dialinecolor}{rgb}{0.000000, 0.000000, 0.000000}
\pgfsetstrokecolor{dialinecolor}
\draw (29.000000\du,14.000000\du)--(29.000000\du,17.500000\du)--(35.000000\du,17.500000\du)--(35.000000\du,14.000000\du)--cycle;
% setfont left to latex
\definecolor{dialinecolor}{rgb}{0.000000, 0.000000, 0.000000}
\pgfsetstrokecolor{dialinecolor}
\node at (32.000000\du,15.840000\du){};
% setfont left to latex
\definecolor{dialinecolor}{rgb}{0.000000, 0.000000, 0.000000}
\pgfsetstrokecolor{dialinecolor}
\node at (31.750000\du,15.000000\du){$\{s_i(t)\}_{i=1}^{\numpow}$};
\node at (31.750000\du,16.640000\du){$\{v_i(t)\}_{i=1}^{\numvol}$};
% setfont left to latex
\definecolor{dialinecolor}{rgb}{0.000000, 0.000000, 0.000000}
\pgfsetstrokecolor{dialinecolor}
\node at (32.000000\du,20.069671\du){Fully};
% setfont left to latex
\definecolor{dialinecolor}{rgb}{0.000000, 0.000000, 0.000000}
\pgfsetstrokecolor{dialinecolor}
\node at (32.000000\du,21.180782\du){observable};
% setfont left to latex
\definecolor{dialinecolor}{rgb}{0.000000, 0.000000, 0.000000}
\pgfsetstrokecolor{dialinecolor}
\node at (32.000000\du,22.291893\du){time series};
% setfont left to latex
\definecolor{dialinecolor}{rgb}{0.000000, 0.000000, 0.000000}
\pgfsetstrokecolor{dialinecolor}
\node at (32.000000\du,11.075739\du){Partially };
% setfont left to latex
\definecolor{dialinecolor}{rgb}{0.000000, 0.000000, 0.000000}
\pgfsetstrokecolor{dialinecolor}
\node at (32.000000\du,12.086850\du){observable};
% setfont left to latex
\definecolor{dialinecolor}{rgb}{0.000000, 0.000000, 0.000000}
\pgfsetstrokecolor{dialinecolor}
\node at (32.000000\du,13.197961\du){time step};
\definecolor{dialinecolor}{rgb}{0.921569, 0.945098, 0.874510}
\pgfsetfillcolor{dialinecolor}
\fill (29.000000\du,23.000000\du)--(29.000000\du,26.100000\du)--(35.000000\du,26.100000\du)--(35.000000\du,23.000000\du)--cycle;
\pgfsetlinewidth{0.100000\du}
\pgfsetdash{}{0pt}
\pgfsetdash{}{0pt}
\pgfsetmiterjoin
\definecolor{dialinecolor}{rgb}{0.000000, 0.000000, 0.000000}
\pgfsetstrokecolor{dialinecolor}
\draw (29.000000\du,23.000000\du)--(29.000000\du,26.100000\du)--(35.000000\du,26.100000\du)--(35.000000\du,23.000000\du)--cycle;
% setfont left to latex
\definecolor{dialinecolor}{rgb}{0.000000, 0.000000, 0.000000}
\pgfsetstrokecolor{dialinecolor}
\node at (32.000000\du,23.7000000\du){$t-T<\tau<t$};
\node at (32.800000\du,24.490000\du){$1\le i\le N$};
% setfont left to latex
\definecolor{dialinecolor}{rgb}{0.000000, 0.000000, 0.000000}
\pgfsetstrokecolor{dialinecolor}
\node at (32.000000\du,25.40000\du){$\{s_i(\tau),v_i(\tau)\}$};
\pgfsetlinewidth{0.100000\du}
\pgfsetdash{}{0pt}
\pgfsetdash{}{0pt}
\pgfsetbuttcap
{
\definecolor{dialinecolor}{rgb}{0.000000, 0.000000, 0.000000}
\pgfsetfillcolor{dialinecolor}
% was here!!!
\pgfsetarrowsend{latex}
\definecolor{dialinecolor}{rgb}{0.000000, 0.000000, 0.000000}
\pgfsetstrokecolor{dialinecolor}
\draw (35.000000\du,24.550000\du)--(39.143750\du,24.500000\du);
}
\definecolor{dialinecolor}{rgb}{0.933333, 0.917647, 0.949020}
\pgfsetfillcolor{dialinecolor}
\fill (39.243750\du,22.000000\du)--(39.243750\du,27.000000\du)--(45.888750\du,27.000000\du)--(45.888750\du,22.000000\du)--cycle;
\pgfsetlinewidth{0.100000\du}
\pgfsetdash{}{0pt}
\pgfsetdash{}{0pt}
\pgfsetmiterjoin
\definecolor{dialinecolor}{rgb}{0.000000, 0.000000, 0.000000}
\pgfsetstrokecolor{dialinecolor}
\draw (39.243750\du,22.000000\du)--(39.243750\du,27.000000\du)--(45.888750\du,27.000000\du)--(45.888750\du,22.000000\du)--cycle;
% setfont left to latex
\definecolor{dialinecolor}{rgb}{0.000000, 0.000000, 0.000000}
\pgfsetstrokecolor{dialinecolor}
\node at (42.616250\du,23.840000\du){LSTM};
% setfont left to latex
\definecolor{dialinecolor}{rgb}{0.000000, 0.000000, 0.000000}
\pgfsetstrokecolor{dialinecolor}
\node at (42.616250\du,24.740000\du){2 layers};
% setfont left to latex
\definecolor{dialinecolor}{rgb}{0.000000, 0.000000, 0.000000}
\pgfsetstrokecolor{dialinecolor}
\node at (42.616250\du,25.540000\du){Tanh activation};
\pgfsetlinewidth{0.100000\du}
\pgfsetdash{}{0pt}
\pgfsetdash{}{0pt}
\pgfsetbuttcap
{
\definecolor{dialinecolor}{rgb}{0.000000, 0.000000, 0.000000}
\pgfsetfillcolor{dialinecolor}
% was here!!!
\pgfsetarrowsend{latex}
\definecolor{dialinecolor}{rgb}{0.000000, 0.000000, 0.000000}
\pgfsetstrokecolor{dialinecolor}
\draw (45.888750\du,24.500000\du)--(48.600000\du,24.500000\du);
}
% setfont left to latex
\definecolor{dialinecolor}{rgb}{0.000000, 0.000000, 0.000000}
\pgfsetstrokecolor{dialinecolor}
\node[anchor=west] at (45.905026\du,23.62\du){$h_{fo}$};
% setfont left to latex
\definecolor{dialinecolor}{rgb}{0.000000, 0.000000, 0.000000}
\pgfsetstrokecolor{dialinecolor}
\node[anchor=west] at (45.905026\du,25.4\du){$N/3$};
% setfont left to latex
\definecolor{dialinecolor}{rgb}{0.000000, 0.000000, 0.000000}
\pgfsetstrokecolor{dialinecolor}
\node[anchor=west] at (35.2500000\du,16.90\du){$2$($\numpow$+$\numvol$)};
% setfont left to latex
\definecolor{dialinecolor}{rgb}{0.000000, 0.000000, 0.000000}
\pgfsetstrokecolor{dialinecolor}
\node[anchor=west] at (49.500000\du,23.050000\du){};
\definecolor{dialinecolor}{rgb}{0.933333, 0.917647, 0.949020}
\pgfsetfillcolor{dialinecolor}
\fill (39.500000\du,14.050000\du)--(39.500000\du,18.000000\du)--(46.100000\du,18.000000\du)--(46.21000000\du,14.050000\du)--cycle;
\pgfsetlinewidth{0.100000\du}
\pgfsetdash{}{0pt}
\pgfsetdash{}{0pt}
\pgfsetmiterjoin
\definecolor{dialinecolor}{rgb}{0.000000, 0.000000, 0.000000}
\pgfsetstrokecolor{dialinecolor}
\draw (39.500000\du,14.050000\du)--(39.500000\du,18.000000\du)--(46.1000000\du,18.000000\du)--(46.1000000\du,14.050000\du)--cycle;
% setfont left to latex
\definecolor{dialinecolor}{rgb}{0.000000, 0.000000, 0.000000}
\pgfsetstrokecolor{dialinecolor}
\node at (42.800000\du,15.465000\du){Fully Connected};
% setfont left to latex
\definecolor{dialinecolor}{rgb}{0.000000, 0.000000, 0.000000}
\pgfsetstrokecolor{dialinecolor}
\node at (42.800000\du,16.265000\du){2 Layers};
% setfont left to latex
\definecolor{dialinecolor}{rgb}{0.000000, 0.000000, 0.000000}
\pgfsetstrokecolor{dialinecolor}
\node at (42.800000\du,17.065000\du){Tanh activations};
% setfont left to latex
\definecolor{dialinecolor}{rgb}{0.000000, 0.000000, 0.000000}
\pgfsetstrokecolor{dialinecolor}
\node[anchor=west] at (46.031803\du,15.079289\du){$h_{po}$};
% setfont left to latex
\definecolor{dialinecolor}{rgb}{0.000000, 0.000000, 0.000000}
\pgfsetstrokecolor{dialinecolor}
\node[anchor=west] at (46.00031803\du,16.979289\du){$N/6$};
\pgfsetlinewidth{0.100000\du}
\pgfsetdash{}{0pt}
\pgfsetdash{}{0pt}
\pgfsetbuttcap
{
\definecolor{dialinecolor}{rgb}{0.000000, 0.000000, 0.000000}
\pgfsetfillcolor{dialinecolor}
% was here!!!
\pgfsetarrowsend{latex}
\definecolor{dialinecolor}{rgb}{0.000000, 0.000000, 0.000000}
\pgfsetstrokecolor{dialinecolor}
\draw (35.000000\du,16.000000\du)--(39.400000\du,16.025000\du);
}
\pgfsetlinewidth{0.100000\du}
\pgfsetdash{}{0pt}
\pgfsetdash{}{0pt}
\pgfsetbuttcap
{
\definecolor{dialinecolor}{rgb}{0.000000, 0.000000, 0.000000}
\pgfsetfillcolor{dialinecolor}
% was here!!!
\pgfsetarrowsend{latex}
\definecolor{dialinecolor}{rgb}{0.000000, 0.000000, 0.000000}
\pgfsetstrokecolor{dialinecolor}
\draw (46.000000\du,16.025000\du)--(48.600000\du,16.000000\du);
}
% setfont left to latex
\definecolor{dialinecolor}{rgb}{0.000000, 0.000000, 0.000000}
\pgfsetstrokecolor{dialinecolor}
\node[anchor=west] at (35.261829\du,25.443934\du){$4N(T$-$1)$};
\definecolor{dialinecolor}{rgb}{0.964706, 0.878431, 0.760784}
\pgfsetfillcolor{dialinecolor}
\fill (52.000000\du,12.000000\du)--(52.000000\du,28.000000\du)--(69.000000\du,28.000000\du)--(69.000000\du,12.000000\du)--cycle;
\pgfsetlinewidth{0.100000\du}
\pgfsetdash{}{0pt}
\pgfsetdash{}{0pt}
\pgfsetmiterjoin
\definecolor{dialinecolor}{rgb}{0.000000, 0.000000, 0.000000}
\pgfsetstrokecolor{dialinecolor}
\draw (52.000000\du,12.000000\du)--(52.000000\du,28.000000\du)--(69.000000\du,28.000000\du)--(69.000000\du,12.000000\du)--cycle;
% setfont left to latex
\definecolor{dialinecolor}{rgb}{0.000000, 0.000000, 0.000000}
\pgfsetstrokecolor{dialinecolor}
\node at (60.500000\du,20.528056\du){};
\pgfsetlinewidth{0.100000\du}
\pgfsetdash{}{0pt}
\pgfsetdash{}{0pt}
\pgfsetbuttcap
{
\definecolor{dialinecolor}{rgb}{0.000000, 0.000000, 0.000000}
\pgfsetfillcolor{dialinecolor}
% was here!!!
\pgfsetarrowsend{latex}
\definecolor{dialinecolor}{rgb}{0.000000, 0.000000, 0.000000}
\pgfsetstrokecolor{dialinecolor}
\draw (68.000000\du,24.500000\du)--(70.648700\du,24.450000\du);
}
\definecolor{dialinecolor}{rgb}{0.921569, 0.945098, 0.874510}
\pgfsetfillcolor{dialinecolor}
% \fill (71.348700\du,21.800000\du)--(71.348700\du,27.100000\du)--(76.381200\du,27.100000\du)--(76.381200\du,21.800000\du)--cycle;
\pgfsetlinewidth{0.100000\du}
\pgfsetdash{}{0pt}
\pgfsetdash{}{0pt}
\pgfsetmiterjoin
\definecolor{dialinecolor}{rgb}{0.000000, 0.000000, 0.000000}
\pgfsetstrokecolor{dialinecolor}
% \draw (71.348700\du,21.800000\du)--(71.348700\du,27.100000\du)--(76.381200\du,27.100000\du)--(76.381200\du,21.800000\du)--cycle;
% setfont left to latex
\definecolor{dialinecolor}{rgb}{0.000000, 0.000000, 0.000000}
\pgfsetstrokecolor{dialinecolor}
\node at (73.864950\du,24.290000\du){};
% setfont left to latex
\definecolor{dialinecolor}{rgb}{0.000000, 0.000000, 0.000000}
\pgfsetstrokecolor{dialinecolor}
\node at (70.8\du,23.600000\du){$\underline{\hat{v}}(t)$};
% setfont left to latex
\definecolor{dialinecolor}{rgb}{0.000000, 0.000000, 0.000000}
\pgfsetstrokecolor{dialinecolor}
\node[anchor=west] at (72.707105\du,21.082842\du){};
% setfont left to latex
\definecolor{dialinecolor}{rgb}{0.000000, 0.000000, 0.000000}
\pgfsetstrokecolor{dialinecolor}
\node[anchor=west] at (57.050000\du,13.100000\du){\LARGE Regressor};
\definecolor{dialinecolor}{rgb}{0.933333, 0.917647, 0.949020}
\pgfsetfillcolor{dialinecolor}
\fill (54.000000\du,22.000000\du)--(54.000000\du,27.000000\du)--(61.547500\du,27.000000\du)--(61.547500\du,22.000000\du)--cycle;
\pgfsetlinewidth{0.100000\du}
\pgfsetdash{}{0pt}
\pgfsetdash{}{0pt}
\pgfsetmiterjoin
\definecolor{dialinecolor}{rgb}{0.000000, 0.000000, 0.000000}
\pgfsetstrokecolor{dialinecolor}
\draw (54.000000\du,22.000000\du)--(54.000000\du,27.000000\du)--(61.547500\du,27.000000\du)--(61.547500\du,22.000000\du)--cycle;
% setfont left to latex
\definecolor{dialinecolor}{rgb}{0.000000, 0.000000, 0.000000}
\pgfsetstrokecolor{dialinecolor}
\node at (57.773750\du,23.940000\du){Fully Connected};
% setfont left to latex
\definecolor{dialinecolor}{rgb}{0.000000, 0.000000, 0.000000}
\pgfsetstrokecolor{dialinecolor} \node at (57.773750\du,24.740000\du){5 Layer};
% setfont left to latex
\definecolor{dialinecolor}{rgb}{0.000000, 0.000000, 0.000000}
\pgfsetstrokecolor{dialinecolor}
\node at (57.773750\du,25.540000\du){Tanh activations};
\pgfsetlinewidth{0.100000\du}
\pgfsetdash{}{0pt}
\pgfsetdash{}{0pt}
\pgfsetbuttcap
{
\definecolor{dialinecolor}{rgb}{0.000000, 0.000000, 0.000000}
\pgfsetfillcolor{dialinecolor}
% was here!!!
\pgfsetarrowsend{latex}
\definecolor{dialinecolor}{rgb}{0.000000, 0.000000, 0.000000}
\pgfsetstrokecolor{dialinecolor}
\draw (49.106066\du,24.469671\du)--(53.944374\du,24.501570\du);
}
% setfont left to latex
\definecolor{dialinecolor}{rgb}{0.000000, 0.000000, 0.000000}
\pgfsetstrokecolor{dialinecolor}
\node[anchor=west] at (62.0\du,25.4\du){$2N$};
\definecolor{dialinecolor}{rgb}{0.933333, 0.917647, 0.949020}
\pgfsetfillcolor{dialinecolor}
\fill (48.500000\du,14.000000\du)--(48.500000\du,27.000000\du)--(49.457500\du,27.000000\du)--(49.457500\du,14.000000\du)--cycle;
\pgfsetlinewidth{0.100000\du}
\pgfsetdash{}{0pt}
\pgfsetdash{}{0pt}
\pgfsetmiterjoin
\definecolor{dialinecolor}{rgb}{0.000000, 0.000000, 0.000000}
\pgfsetstrokecolor{dialinecolor}
\draw (48.500000\du,14.000000\du)--(48.500000\du,27.000000\du)--(49.457500\du,27.000000\du)--(49.457500\du,14.000000\du)--cycle;
% setfont left to latex
\definecolor{dialinecolor}{rgb}{0.000000, 0.000000, 0.000000}
\pgfsetstrokecolor{dialinecolor}
\node at (48.928750\du,16.740000\du){c};
% setfont left to latex
\definecolor{dialinecolor}{rgb}{0.000000, 0.000000, 0.000000}
\pgfsetstrokecolor{dialinecolor}
\node at (48.928750\du,17.540000\du){o};
% setfont left to latex
\definecolor{dialinecolor}{rgb}{0.000000, 0.000000, 0.000000}
\pgfsetstrokecolor{dialinecolor}
\node at (48.928750\du,18.340000\du){n};
% setfont left to latex
\definecolor{dialinecolor}{rgb}{0.000000, 0.000000, 0.000000}
\pgfsetstrokecolor{dialinecolor}
\node at (48.928750\du,19.140000\du){c};
% setfont left to latex
\definecolor{dialinecolor}{rgb}{0.000000, 0.000000, 0.000000}
\pgfsetstrokecolor{dialinecolor}
\node at (48.928750\du,19.940000\du){a};
% setfont left to latex
\definecolor{dialinecolor}{rgb}{0.000000, 0.000000, 0.000000}
\pgfsetstrokecolor{dialinecolor}
\node at (48.928750\du,20.740000\du){t};
% setfont left to latex
\definecolor{dialinecolor}{rgb}{0.000000, 0.000000, 0.000000}
\pgfsetstrokecolor{dialinecolor}
\node at (48.928750\du,21.540000\du){e};
% setfont left to latex
\definecolor{dialinecolor}{rgb}{0.000000, 0.000000, 0.000000}
\pgfsetstrokecolor{dialinecolor}
\node at (48.928750\du,22.340000\du){n};
% setfont left to latex
\definecolor{dialinecolor}{rgb}{0.000000, 0.000000, 0.000000}
\pgfsetstrokecolor{dialinecolor}
\node at (48.928750\du,23.140000\du){a};
% setfont left to latex
\definecolor{dialinecolor}{rgb}{0.000000, 0.000000, 0.000000}
\pgfsetstrokecolor{dialinecolor}
\node at (48.9728750\du,23.940000\du){t};
% setfont left to latex
\definecolor{dialinecolor}{rgb}{0.000000, 0.000000, 0.000000}
\pgfsetstrokecolor{dialinecolor}
\node at (48.928750\du,24.740000\du){e};
\definecolor{dialinecolor}{rgb}{0.933333, 0.917647, 0.949020}
\pgfsetfillcolor{dialinecolor}
\fill (54.000000\du,14.000000\du)--(54.000000\du,18.000000\du)--(61.547500\du,18.000000\du)--(61.547500\du,14.000000\du)--cycle;
\pgfsetlinewidth{0.100000\du}
\pgfsetdash{}{0pt}
\pgfsetdash{}{0pt}
\pgfsetmiterjoin
\definecolor{dialinecolor}{rgb}{0.000000, 0.000000, 0.000000}
\pgfsetstrokecolor{dialinecolor}
\draw (54.000000\du,14.000000\du)--(54.000000\du,18.000000\du)--(61.547500\du,18.000000\du)--(61.547500\du,14.000000\du)--cycle;
% setfont left to latex
\definecolor{dialinecolor}{rgb}{0.000000, 0.000000, 0.000000}
\pgfsetstrokecolor{dialinecolor}
\node at (57.773750\du,15.440000\du){Fully Connected};
% setfont left to latex
\definecolor{dialinecolor}{rgb}{0.000000, 0.000000, 0.000000}
\pgfsetstrokecolor{dialinecolor} \node at (57.773750\du,16.240000\du){1 Layer};
% setfont left to latex
\definecolor{dialinecolor}{rgb}{0.000000, 0.000000, 0.000000}
\pgfsetstrokecolor{dialinecolor}
\node at (57.773750\du,17.040000\du){Tanh activations};
\pgfsetlinewidth{0.100000\du}
\pgfsetdash{}{0pt}
\pgfsetdash{}{0pt}
\pgfsetmiterjoin
\pgfsetbuttcap
{
\definecolor{dialinecolor}{rgb}{0.000000, 0.000000, 0.000000}
\pgfsetfillcolor{dialinecolor}
% was here!!!
\pgfsetarrowsend{latex}
{\pgfsetcornersarced{\pgfpoint{0.000000\du}{0.000000\du}}\definecolor{dialinecolor}{rgb}{0.000000, 0.000000, 0.000000}
\pgfsetstrokecolor{dialinecolor}
\draw (54.000000\du,24.500000\du)--(52.459783\du,24.500000\du)--(52.459783\du,16.000000\du)--(54.000000\du,16.000000\du);
}}
\pgfsetlinewidth{0.100000\du}
\pgfsetdash{}{0pt}
\pgfsetdash{}{0pt}
\pgfsetbuttcap
\pgfsetmiterjoin
\pgfsetlinewidth{0.100000\du}
\pgfsetbuttcap
\pgfsetmiterjoin
\pgfsetdash{}{0pt}
\definecolor{dialinecolor}{rgb}{0.917647, 0.937255, 0.960784}
\pgfsetfillcolor{dialinecolor}
\pgfpathellipse{\pgfpoint{64.500000\du}{24.500000\du}}{\pgfpoint{0.500000\du}{0\du}}{\pgfpoint{0\du}{0.500000\du}}
\pgfusepath{fill}
\definecolor{dialinecolor}{rgb}{0.000000, 0.000000, 0.000000}
\pgfsetstrokecolor{dialinecolor}
\pgfpathellipse{\pgfpoint{64.500000\du}{24.500000\du}}{\pgfpoint{0.500000\du}{0\du}}{\pgfpoint{0\du}{0.500000\du}}
\pgfusepath{stroke}
\pgfsetbuttcap
\pgfsetmiterjoin
\pgfsetdash{}{0pt}
\definecolor{dialinecolor}{rgb}{0.000000, 0.000000, 0.000000}
\pgfsetstrokecolor{dialinecolor}
\draw (64.146450\du,24.146450\du)--(64.853550\du,24.853550\du);
\pgfsetbuttcap
\pgfsetmiterjoin
\pgfsetdash{}{0pt}
\definecolor{dialinecolor}{rgb}{0.000000, 0.000000, 0.000000}
\pgfsetstrokecolor{dialinecolor}
\draw (64.146450\du,24.853550\du)--(64.853550\du,24.146450\du);
\pgfsetlinewidth{0.100000\du}
\pgfsetdash{}{0pt}
\pgfsetdash{}{0pt}
\pgfsetbuttcap
\pgfsetmiterjoin
\pgfsetlinewidth{0.100000\du}
\pgfsetbuttcap
\pgfsetmiterjoin
\pgfsetdash{}{0pt}
\definecolor{dialinecolor}{rgb}{0.917647, 0.937255, 0.960784}
\pgfsetfillcolor{dialinecolor}
\pgfpathellipse{\pgfpoint{67.500000\du}{24.500000\du}}{\pgfpoint{0.500000\du}{0\du}}{\pgfpoint{0\du}{0.500000\du}}
\pgfusepath{fill}
\definecolor{dialinecolor}{rgb}{0.000000, 0.000000, 0.000000}
\pgfsetstrokecolor{dialinecolor}
\pgfpathellipse{\pgfpoint{67.500000\du}{24.500000\du}}{\pgfpoint{0.500000\du}{0\du}}{\pgfpoint{0\du}{0.500000\du}}
\pgfusepath{stroke}
\pgfsetbuttcap
\pgfsetmiterjoin
\pgfsetdash{}{0pt}
\definecolor{dialinecolor}{rgb}{0.000000, 0.000000, 0.000000}
\pgfsetstrokecolor{dialinecolor}
\draw (67.500000\du,24.000000\du)--(67.500000\du,25.000000\du);
\pgfsetbuttcap
\pgfsetmiterjoin
\pgfsetdash{}{0pt}
\definecolor{dialinecolor}{rgb}{0.000000, 0.000000, 0.000000}
\pgfsetstrokecolor{dialinecolor}
\draw (67.000000\du,24.500000\du)--(68.000000\du,24.500000\du);
\pgfsetlinewidth{0.100000\du}
\pgfsetdash{}{0pt}
\pgfsetdash{}{0pt}
\pgfsetbuttcap
{
\definecolor{dialinecolor}{rgb}{0.000000, 0.000000, 0.000000}
\pgfsetfillcolor{dialinecolor}
% was here!!!
\pgfsetarrowsend{latex}
\definecolor{dialinecolor}{rgb}{0.000000, 0.000000, 0.000000}
\pgfsetstrokecolor{dialinecolor}
\draw (61.547500\du,24.500000\du)--(64.000000\du,24.500000\du);
}
\definecolor{dialinecolor}{rgb}{0.933333, 0.917647, 0.949020}
\pgfsetfillcolor{dialinecolor}
% \fill (64.000000\du,16.000000\du)--(64.000000\du,18.000000\du)--(65.000000\du,18.000000\du)--(65.000000\du,16.000000\du)--cycle;
\pgfsetlinewidth{0.100000\du}
\pgfsetdash{}{0pt}
\pgfsetdash{}{0pt}
\pgfsetmiterjoin
\definecolor{dialinecolor}{rgb}{0.000000, 0.000000, 0.000000}
\pgfsetstrokecolor{dialinecolor}
% \draw (64.000000\du,16.000000\du)--(64.000000\du,18.000000\du)--(65.000000\du,18.000000\du)--(65.000000\du,16.000000\du)--cycle;
% setfont left to latex
\definecolor{dialinecolor}{rgb}{0.000000, 0.000000, 0.000000}
\pgfsetstrokecolor{dialinecolor}
\node at (62.500000\du,16.50000\du){$w$};
\definecolor{dialinecolor}{rgb}{0.933333, 0.917647, 0.949020}
\pgfsetfillcolor{dialinecolor}
% \fill (64.000000\du,14.000000\du)--(64.000000\du,16.000000\du)--(65.000000\du,16.000000\du)--(65.000000\du,14.000000\du)--cycle;
\pgfsetlinewidth{0.100000\du}
\pgfsetdash{}{0pt}
\pgfsetdash{}{0pt}
\pgfsetmiterjoin
\definecolor{dialinecolor}{rgb}{0.000000, 0.000000, 0.000000}
\pgfsetstrokecolor{dialinecolor}
% \draw (64.000000\du,14.000000\du)--(64.000000\du,16.000000\du)--(65.000000\du,16.000000\du)--(65.000000\du,14.000000\du)--cycle;
% setfont left to latex
\definecolor{dialinecolor}{rgb}{0.000000, 0.000000, 0.000000}
\pgfsetstrokecolor{dialinecolor}
\node at (66.500000\du,14.400000\du){$b$};
\pgfsetlinewidth{0.100000\du}
\pgfsetdash{}{0pt}
\pgfsetdash{}{0pt}
\pgfsetbuttcap
% {
% \definecolor{dialinecolor}{rgb}{0.000000, 0.000000, 0.000000}
% \pgfsetfillcolor{dialinecolor}
% % was here!!!
% \pgfsetarrowsend{latex}
% \definecolor{dialinecolor}{rgb}{0.000000, 0.000000, 0.000000}
% \pgfsetstrokecolor{dialinecolor}
% \draw (61.547500\du,15.000000\du)--(64.000000\du,15.000000\du);
% }
\pgfsetlinewidth{0.100000\du}
\pgfsetdash{}{0pt}
\pgfsetdash{}{0pt}
\pgfsetbuttcap
% {
% \definecolor{dialinecolor}{rgb}{0.000000, 0.000000, 0.000000}
% \pgfsetfillcolor{dialinecolor}
% % was here!!!
% \pgfsetarrowsend{latex}
% \definecolor{dialinecolor}{rgb}{0.000000, 0.000000, 0.000000}
% \pgfsetstrokecolor{dialinecolor}
% \draw (61.547500\du,17.000000\du)--(64.000000\du,17.000000\du);
% }
\pgfsetlinewidth{0.100000\du}
\pgfsetdash{}{0pt}
\pgfsetdash{}{0pt}
\pgfsetmiterjoin
\pgfsetbuttcap
{
\definecolor{dialinecolor}{rgb}{0.000000, 0.000000, 0.000000}
\pgfsetfillcolor{dialinecolor}
% was here!!!
\pgfsetarrowsend{latex}
{\pgfsetcornersarced{\pgfpoint{0.000000\du}{0.000000\du}}\definecolor{dialinecolor}{rgb}{0.000000, 0.000000, 0.000000}
\pgfsetstrokecolor{dialinecolor}
\draw (61.500000\du,15.000000\du)--(67.500000\du,15.000000\du)--(67.500000\du,24.000000\du);
}}
\pgfsetlinewidth{0.100000\du}
\pgfsetdash{}{0pt}
\pgfsetdash{}{0pt}
\pgfsetbuttcap
{
\definecolor{dialinecolor}{rgb}{0.000000, 0.000000, 0.000000}
\pgfsetfillcolor{dialinecolor}
% was here!!!
\pgfsetarrowsend{latex}
\definecolor{dialinecolor}{rgb}{0.000000, 0.000000, 0.000000}
\pgfsetstrokecolor{dialinecolor}
\draw (61.50000\du,17.0000000\du)--(64.500000\du,17.000000\du)--(64.500000\du,24.000000\du);
}
\pgfsetlinewidth{0.100000\du}
\pgfsetdash{}{0pt}
\pgfsetdash{}{0pt}
\pgfsetbuttcap
{
\definecolor{dialinecolor}{rgb}{0.000000, 0.000000, 0.000000}
\pgfsetfillcolor{dialinecolor}
% was here!!!
\pgfsetarrowsend{latex}
\definecolor{dialinecolor}{rgb}{0.000000, 0.000000, 0.000000}
\pgfsetstrokecolor{dialinecolor}
\draw (65.000000\du,24.500000\du)--(67.000000\du,24.500000\du);
}
% setfont left to latex
\definecolor{dialinecolor}{rgb}{0.000000, 0.000000, 0.000000}
\pgfsetstrokecolor{dialinecolor}
\node[anchor=west] at (65.57158\du,15.555026\du){$2N$};
% setfont left to latex
\definecolor{dialinecolor}{rgb}{0.000000, 0.000000, 0.000000}
\pgfsetstrokecolor{dialinecolor}
\node[anchor=west] at (61.5\du,17.50000\du){$2N$};
% setfont left to latex
\definecolor{dialinecolor}{rgb}{0.000000, 0.000000, 0.000000}
\pgfsetstrokecolor{dialinecolor}
\node[anchor=west] at (65.255026\du,25.4\du){$2N$};
% setfont left to latex
\definecolor{dialinecolor}{rgb}{0.000000, 0.000000, 0.000000}
\pgfsetstrokecolor{dialinecolor}
\node[anchor=west] at (67.255026\du,25.4\du){$2N$};
% setfont left to latex
\definecolor{dialinecolor}{rgb}{0.000000, 0.000000, 0.000000}
\pgfsetstrokecolor{dialinecolor}
\node[anchor=west] at (36.563906\du,13.000000\du){\LARGE Feature Extractor};
% setfont left to latex
\definecolor{dialinecolor}{rgb}{0.000000, 0.000000, 0.000000}
\pgfsetstrokecolor{dialinecolor}
\node[anchor=west] at (50.283580\du,23.8\du){$h$};
% setfont left to latex
\definecolor{dialinecolor}{rgb}{0.000000, 0.000000, 0.000000}
\pgfsetstrokecolor{dialinecolor}
\node[anchor=west] at (49.753580\du,25.4\du){$N/2$};
\end{tikzpicture}

%% file: manuscript.bbl
% Generated by IEEEtran.bst, version: 1.14 (2015/08/26)
\begin{thebibliography}{10}
\providecommand{\url}[1]{#1}
\csname url@samestyle\endcsname
\providecommand{\newblock}{\relax}
\providecommand{\bibinfo}[2]{#2}
\providecommand{\BIBentrySTDinterwordspacing}{\spaceskip=0pt\relax}
\providecommand{\BIBentryALTinterwordstretchfactor}{4}
\providecommand{\BIBentryALTinterwordspacing}{\spaceskip=\fontdimen2\font plus
\BIBentryALTinterwordstretchfactor\fontdimen3\font minus
  \fontdimen4\font\relax}
\providecommand{\BIBforeignlanguage}[2]{{%
\expandafter\ifx\csname l@#1\endcsname\relax
\typeout{** WARNING: IEEEtran.bst: No hyphenation pattern has been}%
\typeout{** loaded for the language `#1'. Using the pattern for}%
\typeout{** the default language instead.}%
\else
\language=\csname l@#1\endcsname
\fi
#2}}
\providecommand{\BIBdecl}{\relax}
\BIBdecl

\bibitem{Abur2004}
A.~Abur and A.~G. Exposito, \emph{{Power System State Estimation: Theory and
  Implementation}}.\hskip 1em plus 0.5em minus 0.4em\relax Abingdon: Dekker,
  2004.

\bibitem{baran2001challenges}
M.~E. {Baran}, ``Challenges in state estimation on distribution systems,'' in
  \emph{2001 Power Engineering Society Summer Meeting. Conference Proceedings
  (Cat. No.01CH37262)}, vol.~1, July 2001, pp. 429--433.

\bibitem{soltan2018react}
S.~Soltan, M.~Yannakakis, and G.~Zussman, ``{REACT} to cyber attacks on power
  grids,'' \emph{IEEE Transactions on Network Science and Engineering}, vol.~6,
  no.~3, pp. 459--473, 2018.

\bibitem{soltan2019expose}
S.~{Soltan} and G.~{Zussman}, ``{EXPOSE} the line failures following a
  cyber-physical attack on the power grid,'' \emph{IEEE Transactions on Control
  of Network Systems}, vol.~6, no.~1, pp. 451--461, 2019.

\bibitem{soltan2016quantifying}
S.~Soltan and G.~Zussman, ``Quantifying the effect of k-line failures in power
  grids,'' in \emph{IEEE Power and Energy Society General Meeting (PESGM)},
  July 2016.

\bibitem{ostrometzky2019irradiance}
J.~Ostrometzky, A.~Bernstein, and G.~Zussman, ``Irradiance field reconstruction
  from partial observability of solar radiation,'' \emph{To appear in IEEE
  Geoscience and Remote Sensing Letters}, 2019.

\bibitem{primadianto2017review}
A.~{Primadianto} and C.~{Lu}, ``A review on distribution system state
  estimation,'' \emph{IEEE Transactions on Power Systems}, vol.~32, no.~5, pp.
  3875--3883, Sept. 2017.

\bibitem{wang2019distribution}
G.~Wang, G.~B. Giannakis, J.~Chen, and J.~Sun, ``Distribution system state
  estimation: an overview of recent developments,'' \emph{Frontiers of
  Information Technology {\&} Electronic Engineering}, vol.~20, no.~1, pp.
  4--17, Jan. 2019.

\bibitem{singh2009}
R.~Singh, B.~C. Pal, and R.~B. Vinter, ``Measurement placement in distribution
  system state estimation,'' \emph{IEEE Transactions on Power Systems},
  vol.~24, no.~2, pp. 668--675, 2009.

\bibitem{bhela2018}
S.~Bhela, V.~Kekatos, and S.~Veeramachaneni, ``Enhancing observability in
  distribution grids using smart meter data,'' \emph{IEEE Transactions on Smart
  Grid}, vol.~9, no.~6, pp. 5953--5961, 2018.

\bibitem{YCNN}
H.~Jiang and Y.~Zhang, ``Short-term distribution system state forecast based on
  optimal synchrophasor sensor placement and extreme learning machine,'' in
  \emph{IEEE Power and Energy Society General Meeting (PESGM)}, July 2016.

\bibitem{Clements2011}
K.~A. {Clements}, ``The impact of pseudo-measurements on state estimator
  accuracy,'' in \emph{IEEE Power and Energy Society General Meeting (PESGM)},
  July 2011.

\bibitem{manitsas2012}
E.~Manitsas, R.~Singh, B.~C. Pal, and G.~Strbac, ``Distribution system state
  estimation using an artificial neural network approach for pseudo measurement
  modeling,'' \emph{IEEE Transactions on Power Systems}, vol.~27, no.~4, pp.
  1888--1896, Nov 2012.

\bibitem{wu2013}
J.~Wu, Y.~He, and N.~Jenkins, ``A robust state estimator for medium voltage
  distribution networks,'' \emph{IEEE Transactions on Power Systems}, vol.~28,
  no.~2, pp. 1008--1016, 2013.

\bibitem{gao_wang2016}
P.~Gao, M.~Wang, S.~G. Ghiocel, J.~H. Chow, B.~Fardanesh, and G.~Stefopoulos,
  ``Missing data recovery by exploiting low-dimensionality in power system
  synchrophasor measurements,'' \emph{{IEEE} Transactions on Power Systems},
  vol.~31, no.~2, pp. 1006--1013, 2016.

\bibitem{Genes2019}
C.~{Genes}, I.~{Esnaola}, S.~M. {Perlaza}, L.~F. {Ochoa}, and D.~{Coca},
  ``Robust recovery of missing data in electricity distribution systems,''
  \emph{IEEE Transactions on Smart Grid}, vol.~10, no.~4, pp. 4057--4067, 2019.

\bibitem{Liao2019}
M.~{Liao}, D.~{Shi}, Z.~{Yu}, Z.~{Yi}, Z.~{Wang}, and Y.~{Xiang}, ``An
  alternating direction method of multipliers based approach for pmu data
  recovery,'' \emph{IEEE Transactions on Smart Grid}, vol.~10, no.~4, pp.
  4554--4565, 2019.

\bibitem{Schmitt}
P.~L. {Donti}, L.~Yajing, A.~J. Schmitt, A.~{Bernstein}, Y.~Rui, and
  Y.~{Zhang}, ``Matrix completion for low-observability voltage estimation,''
  \emph{arXiv preprint arXiv:1801.09799}, 2018.

\bibitem{zamzam2019physics}
A.~S. Zamzam and N.~D. Sidiropoulos, ``Physics-aware neural networks for
  distribution system state estimation,'' \emph{arXiv preprint
  arXiv:1903.09669}, 2019.

\bibitem{schneider2018ieee37}
K.~P. {Schneider}, B.~A. {Mather}, B.~C. {Pal}, C.~. {Ten}, G.~J. {Shirek},
  H.~{Zhu}, J.~C. {Fuller}, J.~L.~R. {Pereira}, L.~F. {Ochoa}, L.~R. {de
  Araujo}, R.~C. {Dugan}, S.~{Matthias}, S.~{Paudyal}, T.~E. {McDermott}, and
  W.~{Kersting}, ``Analytic considerations and design basis for the ieee
  distribution test feeders,'' \emph{IEEE Transactions on Power Systems},
  vol.~33, no.~3, pp. 3181--3188, 2018.

\bibitem{linearLoadFlow}
A.~Bernstein, C.~Wang, E.~Dallanese, J.-Y.~L. Boudec, and C.~Zhao, ``Load-flow
  in multiphase distribution networks: Existence, uniqueness, non-singularity,
  and linear models,'' \emph{IEEE Transactions on Power Systems}, pp. 1--1,
  2017.

\bibitem{gers1999learning}
J.~S. F.A.~Gers and F.~Cummins, ``\BIBforeignlanguage{English}{Learning to
  forget: continual prediction with lstm},''
  \emph{\BIBforeignlanguage{English}{IET Conference Proceedings}}, pp.
  850--855, 1999.

\bibitem{nielsen2015neural}
M.~A. Nielsen, \emph{Neural networks and deep learning}.\hskip 1em plus 0.5em
  minus 0.4em\relax Determination press San Francisco, CA, USA, 2015, vol.
  2018.

\bibitem{bank2013development}
J.~Bank and J.~Hambrick, ``Development of a high resolution, real time,
  distribution-level metering system and associated visualization, modeling,
  and data analysis functions,'' National Renewable Energy Lab.(NREL), Golden,
  CO (United States), NREL/TP-5500-56610,, Tech. Rep., 2013.

\bibitem{oppenheim1999discrete}
A.~V. Oppenheim, \emph{Discrete-time signal processing}.\hskip 1em plus 0.5em
  minus 0.4em\relax Pearson Education India, 1999.

\bibitem{zimmerman2010matpower}
R.~D. Zimmerman, C.~E. Murillo-S{\'a}nchez, and R.~J. Thomas, ``Matpower:
  Steady-state operations, planning, and analysis tools for power systems
  research and education,'' \emph{IEEE Transactions on power systems}, vol.~26,
  no.~1, pp. 12--19, 2010.

\bibitem{kingma2014adam}
D.~P. Kingma and J.~Ba, ``Adam: {A} method for stochastic optimization,'' in
  \emph{3rd International Conference on Learning Representations ({ICLR})},
  2015.

\bibitem{trabelsi2017deep}
C.~Trabelsi, O.~Bilaniuk, Y.~Zhang, D.~Serdyuk, S.~Subramanian, J.~F. Santos,
  S.~Mehri, N.~Rostamzadeh, Y.~Bengio, and C.~J. Pal, ``Deep complex
  networks,'' \emph{arXiv preprint arXiv:1705.09792}, 2017.

\end{thebibliography}
